\documentclass[11pt]{article}

\usepackage[usenames,dvipsnames]{pstricks}
\usepackage{epsfig}
\usepackage{pst-grad} % For gradients
\usepackage[space]{grffile} % For spaces in paths
\usepackage{etoolbox} % For spaces in paths
\makeatletter % For spaces in paths
\patchcmd\Gread@eps{\@inputcheck#1 }{\@inputcheck"#1"\relax}{}{}
\makeatother

\usepackage{graphicx}
\usepackage{amsmath}
\usepackage{psfrag}
\usepackage{color,soul}
\usepackage{natbib}
\usepackage{siunitx}
\usepackage{float}
\usepackage{tikz}
\usepackage{multirow}
\usepackage{bm}
\usepackage{textgreek}
\usepackage{url}
\usepackage{amssymb}
\usepackage{wasysym}
\usepackage{xfrac}
\usepackage{setspace}
\usepackage{gensymb}
\usepackage{mathtools}
\usepackage{arydshln}
\usepackage{hyperref}
\usepackage{cleveref}
\usepackage{caption}
\usepackage{subcaption}
\usepackage[percent]{overpic}

\DeclareMathOperator{\Tr}{Tr}

\definecolor{myred}{rgb}{0.5, 0, 0.0}
\definecolor{myblue}{rgb}{0, 0, 0.5}

\usepackage[noblocks]{authblk}
\usepackage[margin=1in]{geometry}
\usepackage[title]{appendix}

\newcommand\affiliation[1]{\gdef\@affiliation{\let\aff\aff@inst#1}}
\gdef\@affiliation{}
\def\email#1{Email address for correspondence: #1}
\def\aff#1{\ignorespaces\textsuperscript{#1}}
\def\corresp#1{\unskip\thanks{#1}}
\numberwithin{equation}{section}

\renewenvironment{abstract}
{\begin{quote}
\noindent \rule{\linewidth}{.5pt}\par{\bfseries \abstractname.}}
{\medskip\noindent \rule{\linewidth}{.5pt}
\end{quote}
}

  \DeclareTextFontCommand\textsfi{\usefont{OT1}{cmss}{m}{sl}}
  \DeclareMathAlphabet\mathsfi            {OT1}{cmss}{m}{sl}
  \DeclareTextFontCommand\textsfb{\usefont{OT1}{cmss}{bx}{n}}
  \DeclareMathAlphabet\mathsfb            {OT1}{cmss}{bx}{n}
  \DeclareTextFontCommand\textsfbi{\usefont{OT1}{cmss}{m}{sl}}
  \DeclareMathAlphabet\mathsfbi            {OT1}{cmss}{m}{sl}
  \IfFileExists{t1phv.fd}
  {\DeclareTextFontCommand\textsfbi{\usefont{T1}{phv}{b}{it}}
  \DeclareMathAlphabet\mathsfbi            {T1}{phv}{b}{it}
  }{}
  \IfFileExists{ot1phv.fd}
  {\DeclareTextFontCommand\textsfbi{\usefont{OT1}{phv}{b}{it}}
  \DeclareMathAlphabet\mathsfbi            {OT1}{phv}{b}{it}
  }{}
  
\DeclareSymbolFont{matha}{OML}{txmi}{m}{it}% txfonts

\crefformat{section}{\S#2#1#3} % see manual of cleveref, section 8.2.1
\crefformat{subsection}{\S#2#1#3}
\crefformat{subsubsection}{\S#2#1#3}

\definecolor{darkblue}{rgb}{0,0,0.80}

\hypersetup{
    colorlinks=true,
    linkcolor={darkblue},
    citecolor={darkblue},
    filecolor=black,
    urlcolor={darkblue},
    plainpages=false,
}

\title{\bf Paper title}

\title{\bf Beyond optimal disturbances: a statistical framework for transient growth}

\author[1]{\bf Peter Frame\corresp{\email{pframe@umich.edu}}}
\author[1]{\bf Aaron Towne}

\affil[1]{\normalsize Department of Mechanical Engineering, University of Michigan, Ann Arbor, MI, USA \vspace{-1cm}}

\date{}
\begin{document}
\maketitle

% Abstract
\begin{abstract}
The theory of transient growth describes how linear mechanisms can cause temporary amplification of disturbances even when the linearized system is asymptotically stable as defined by its eigenvalues. This growth is traditionally quantified by finding the initial disturbance that generates the maximum response, in terms of energy gain, at the peak time of its evolution. While this bounds the growth, it can vastly overstate the growth of a real disturbance. In this paper, we introduce a statistical perspective on transient growth that models statistics of the energy amplification of the disturbances. We derive a formula for the mean energy amplification and two-point spatial correlation of the growing disturbance as a function of the two-point spatial correlation of the initial disturbance. The eigendecomposition of the correlation provides the most prevalent structures, which are the statistical analog of the standard left singular vectors of the evolution operator. We also derive an accurate approximation of the probability density function of the energy of the growing disturbance, from which confidence bounds on the growth can be obtained. Applying our analysis to Poisseuille flow yields a number of observations. First, the mean gain can be drastically smaller than the maximum, especially when the disturbances are broadband in wavenumber content. In these cases, it is exceedingly unlikely to achieve near-optimal growth due to the exponential behavior which we observe in the probability density function. Second, the characteristic length scale of the initial disturbances, encapsulated by the spatial decay within the initial correlation function, has a significant impact on the expected growth; specifically, large-scale initial disturbances produce orders-of-magnitude-larger expected growth than smaller scales, indicating that the length scale of incoming disturbances may be key in determining whether transient growth leads to transition for a particular flow. Finally, while the optimal growth scales quadratically with Reynolds number, we observe that the mean energy amplification scales only linearly for certain reasonable choices of the initial correlations. \\  % DO NOT DELETE THE \\
\end{abstract}

%%%%%%%%%%%%%%%%%%%%%%%%%%%%%%%%%%%%%%%%%%%%%%%%%%%%%%%%%%%%%%%%%%%%%%%%%%%%%%%
% -----------------------------------------------------------------------------
% -- Introduction
% -----------------------------------------------------------------------------
\section{Introduction}
\label{Sec:Intro}
A natural approach for analyzing the stability of a steady fluid flow is to linearize and calculate the eigenvalues of the linearized Navier-Stokes operator. Underlying this analysis is the assumption that there will be disturbances to the steady flow, and though their magnitude is difficult to know a priori, it is likely a value small enough that nonlinear mechanisms are not relevant. This approach, known as modal stability theory, is agnostic to the shape of any particular disturbance --- if there is a positive eigenvalue, any disturbance will grow, otherwise, any disturbance will decay asymptotically. However, the modal approach predicts stability when experiments tell us otherwise. Famously, Reynolds found that at high velocities, pipe flow transitions to turbulence \cite{Reynolds1883}. Efforts to ground this instability in modal theory floundered: pipe flow has all stable eigenvalues. The same is true for Couette flow as well as plane Poisseuille flow at low Reynolds numbers; these flows have only stable eigenvalues, but are observed to transition \cite{tillmark_alfredsson_1992}. 
\\

The key to their instability can be, in fact, a linear mechanism \cite{Schmid07}. Perhaps counterintuitively, a linearized Navier-Stokes operator with all stable eigenvalues can lead to short-term growth in the magnitude of disturbances before they decay at the rate prescribed by the least stable eigenvalue. This transient growth is possible only when the linearized Navier-Stokes operator is non-normal, i.e., its eigenvectors are not orthogonal. This permits one eigenvector to initially subtract from another, but this cancellation can cease if one eigenvector vanishes faster than the other, leading to growth. The magnitude of this growth can be remarkable --- often more than one-thousand-fold at its peak \cite{Trefethen91}. While the initial disturbances are assumed to be too small for nonlinear effects to be important, when they are amplified by three orders of magnitude, the assumption of linearity may no longer hold, and nonlinearities may bring the flow away from the laminar steady state. Rather than leading directly to transition, the nonlinearities activated by the amplified disturbance might bring the flow to a new state. Instability in this state, known as secondary instability, is more likely than the primary growth to lead to turbulence \cite{SH}.
\\

The metric reported in the literature to quantify transient growth is the ratio of kinetic energy of the maximally amplified disturbance to its initial kinetic energy. This metric is usually referred to as $G(t)$, though in this paper we call it $G^{opt}(t)$ to distinguish it from suboptimal and mean growths. Significant effort has been devoted to studying $G^{opt}(t)$ both analytically and numerically. In channel flow, it can be shown to have quadratic dependence on the Reynolds number $Re$ when the product of the streamwise wavenumber and Reynolds number is small, $\alpha Re << 1$ \cite{Gustavsson91}. Under the same conditions, the time at which the maximum occurs increases linearly with $Re$. Indeed, numerical experiments show that there is quadratic scaling in the optimal growth and linear scaling in the optimal time for plane-Poisseuille \cite{Trefethen91}, Couette \cite{Trefethen91}, Blasius boundary layer \cite{Butler92, Hanifi96}, and pipe \cite{SH94} flows. In all of these cases, the optimal streamwise wavenumber $\alpha$ is zero or very small, and the optimal spanwise wavenumber $\beta$ is order unity \cite{SH}. 
\\

Minimal seed theory \cite{Kerswell18} provides a nonlinear analog of optimal transient growth analysis. At each initial energy, it identifies the disturbance that achieves the greatest growth when evolved according to the full nonlinear Navier-Stokes equations. When the initial energy is just large enough that the optimal disturbance leads to sustained turbulence, the disturbance is called the minimal seed --- the smallest disturbance leading to transition. Though minimal seeds are initially amplified by linear mechanisms \cite{Pringle12}, they can differ substantially in shape from the optimal disturbances in linear transient growth \cite{Pringle10}. This gives a lower bound for the energy level that disturbances must achieve to spark transition.
\\

In either the linear or nonlinear context, considering optimal disturbances gives an upper bound on the growth experienced by disturbances, but we propose that a more complete picture of the possible growth is needed. In linear transient growth, only the optimal initial disturbance experiences $G^{opt}$ growth. Indeed, if the initial disturbance were one of the eigenvectors of the linearized Navier-Stokes operator, it would decay monotonically. Of course, real disturbances to the flow will not exactly coincide with the optimally amplified disturbance, so in order to quantify their growth, one needs to explore the space of suboptimal disturbances. Is most of this space inhabited by disturbances that decay or by ones that grow? Is the growth of real disturbances on the order of $G^{opt}$, on average? What is the probability that a random disturbance will come close to $G^{opt}$?
\\

Motivated by these questions, we investigate transient growth from a statistical perspective in this paper. A statistical view serves both to model the uncertainty and variation in the spatial form of initial disturbances and to fully explore the high dimensional space that these disturbances occupy. We derive an equation for the mean energy of the amplified random disturbances, and dividing this by the mean initial energy gives a metric for the mean energy amplification, which we term $G^{mean}$. This depends on the statistics of the incoming disturbances, and the formula we report for $G^{mean}$ involves the correlation matrix of the initial disturbances. The correlation matrix at time $t$ can also be derived in terms of the initial correlations. Its eigendecomposition can be viewed as a particular variant of proper orthogonal decomposition and provides the most statistically prevalent structures, which serve as the statistical analog of the left singular vectors of the evolution matrix.
\\

Quantifying the likelihood that a disturbance grows beyond a particular level requires knowledge of the probability density function (PDF) of the energy amplification. Whereas the mean energy amplification depends only on the correlation matrix of the incoming disturbances, the entire distribution of incoming disturbances is needed to calculate the PDF of the growth. Moreover, there is no general formula relating the two. However, we observe empirically that the PDF is nearly exponential, and this leads to an approximation strategy for it. We use the approximate PDF to derive accurate confidence bounds on the growth, i.e., energy levels which $p\%$ of the disturbances do not exceed, for some desired $p$. The exponential behavior of the PDF also means that if $G^{mean}$ is significantly below $G^{opt}$, it is extremely unlikely for an initial disturbance to achieve near-$G^{opt}$ growth.
\\

Throughout the paper, we demonstrate the statistical framework using plane-Poisseuille flow. Equipped with a statistical lens, numerous observations readily emerge. At each wavenumber pair $(\alpha,\beta)$, the correlation length in the wall-normal direction has a dramatic impact on $G^{mean}$, with correlation lengths on the order of the channel half-height growing to nearly half of $G^{opt}$. If, however, the correlation length is short compared to the channel half-height, $G^{mean}$ can be orders of magnitude smaller than $G^{opt}$. $G^{opt}$ and $G^{mean}$ achieve their maximum values at similar locations in wavenumber space, but the peak is substantially narrower in $\alpha$ for $G^{mean}$. This indicates that 3-dimensional disturbances, ones which contain a range of wavenumbers, further undershoot $G^{opt}$. In the three-dimensional case, $G^{mean}$ is a function of the three-dimensional correlation matrix. We observe that when this correlation is isotropic, $G^{mean}$ is roughly $2\%$ of $G^{opt}$ at $Re = 1000$. Surprisingly, we find that $G^{mean}$ scales nearly linearly with $Re$, so the gap between it and $G^{opt}$ widens with increasing Reynolds number. Therefore, $G^{opt}$ increasingly overstates the growth of random disturbances.
\\

Even considering disturbances near the optimal wavenumber pair ($\alpha = 0$, $\beta = 2$), the probability of exceeding certain levels of growth can be extremely low. We show that the distribution of energy is nearly exponential, i.e., the probability of exceeding a particular energy level decays exponentially. Therefore, if $G^{mean}$ is relatively small relative to $G^{opt}$, there is little chance of observing growth on the order of the optimal value. For a correlation length of one-fourth the channel half-height, fewer than $0.01 \%$ of disturbances achieve $G^{opt}/2$ growth for $Re = 1000$. 
\\

The combined effects of the non-normality of the linearized Navier-Stokes operator and randomness have been analyzed before. In particular, Farrell and Ioannou \cite{Farrell93} considered the linearized Navier-Stokes equations forced continuously by white-in-time noise with some spatial correlation. They showed that the expected energy, once statistical stationarity is reached, can be obtained by solving a Lyapunov equation involving the linearized Navier-Stokes operator. Our study is distinct from this work in that we instead consider the physical model of transient growth --- impulsive disturbances that evolve unforced under the action of the linearized Navier-Stokes equations. This leads to a dependence on the length scales present in the initial disturbances, which we investigate extensively. The present work is also different from what has been called statistical stability \cite{Malkus56,Markeviciute22}. That work is concerned with the stability of the statistical state of turbulent flow, whereas our study investigates statistics of transient growth.
\\

The remainder of the paper is organized as follows. In \S~\ref{sec:flows}, plane-Poisseuille flow and the numerics used to perform the calculations are described. \S~\ref{sec:Transient_Growth} gives a review of transient growth. In \S~\ref{sec:exp_amplification}, we derive a formula for the mean energy amplification and compare it to the optimal growth for Poisseuille flow, first for disturbances at one pair of wavenumbers, then for disturbances containing a range of wavenumbers. We investigate the probability density function of the growth and detail an accurate approximation strategy for it in \S~\ref{sec:Estimate_PDF}. Finally, in \S~\ref{sec:Conclusions}, we conclude the paper and offer some closing remarks.
\\

%%%%%%%%%%%%%%%%%%%%%%%%%%%%%%%%%%%%%%%%%%%%%%%%%%%%%%%%%%%%%%%%%%%%%%%%%%%%%%%
% -----------------------------------------------------------------------------
% -- Next 
% -----------------------------------------------------------------------------

\section{Flow description and numerics} \label{sec:flows}
Plane-Poissseuille flow is the steady, laminar flow between two plates separated by $2h$ in the $y$ direction. The flow is in the $x$ direction, and the plates are infinite in both the streamwise ($x$) and spanwise ($z$) directions. It is driven by a constant pressure gradient and the streamwise velocity field is given by 
\begin{equation}
    U(y) = -\frac{1}{2\mu} \frac{d p}{d x}(h^2 - y^2) \text{.}
\end{equation}
The flow is then non-dimensionalized by the channel half-height and the centerline velocity. Because the governing equations and base-flow are homogenous in $x$ and $z$, it is convenient to take the Fourier transform of disturbances to the base flow in these directions. The associated wavenumbers in the streamwise and spanwise directions are denoted $\alpha$ and $\beta$, respectively. For example, the transformed wall-normal velocity is 
\begin{equation}
     \hat{v}(y,\alpha,\beta) = \int_{-\infty}^{\infty} \int_{-\infty}^{\infty} v(x,y,z) e^{-i(\alpha x + \beta z)}dx dz \text{.}   
\end{equation}
Employing the usual velocity-vorticity formulation of the linearized Navier-Stokes equations yields the following equations for the evolution of disturbances \cite{Reddy93},

\begin{equation} \label{Flow:LNS}
    \frac{\partial}{\partial t} \begin{bmatrix}
        \hat{v} \\
        \hat{\eta}
    \end{bmatrix}
     =     -i \begin{bmatrix}
        \mathcal{L}_{OS} && 0 \\
        \mathcal{L}_{C} && \mathcal{L}_{SQ}
    \end{bmatrix}
    \begin{bmatrix}
        \hat{v} \\
        \hat{\eta}
    \end{bmatrix} \text{.}
\end{equation} 
The Orr-Sommerfeld, cross-term, and Squire operators are
\begin{subequations}
\begin{equation}
    \mathcal{L}_{OS} = - \left(\frac{\partial^2}{\partial y^2} - k^2 \right) ^{-1} \left[ \frac{1}{iRe} \left( \frac{\partial^2}{\partial y^2} - k^2 \right)^2 - \alpha U \left(\frac{\partial^2}{\partial y^2} - k^2 \right) + \alpha U'' \right] \text{,}
\end{equation}
\begin{equation}
    \mathcal{L}_{C} = \beta U' \text{,}
\end{equation}
\begin{equation}
    \mathcal{L}_{SQ} = \alpha U -  \frac{1}{iRe} \left( \frac{\partial^2}{\partial y^2} - k^2 \right)\text{.}
\end{equation}
\end{subequations}
Above, all quantities are non-dimensionalized, and $U = U(y)$ is the base-flow, $\hat{\eta}$ is the transformed wall-normal vorticity, $k^2 = \alpha^2 + \beta^2$ is the squared wavevector magnitude, and $(\cdot)^{'}$ indicates a wall-normal derivative $\frac{\partial}{\partial y}$. We use the code provided in \cite{SH}, which uses a Chebyshev discretization of the linearized Navier-Stokes equations (\ref{Flow:LNS}) \cite{Herbert77,Reddy93}. All norms presented in our numerical results are based on the kinetic energy of a disturbance. It can be shown, by using incompressibility and Parseval's theorem, that the energy of a disturbance in the transformed velocity-vorticity coordinates is \cite{Gustavsson86}
\begin{equation}
    E = \int_{-\infty}^{\infty}\int_{-\infty}^{\infty} \frac{1}{2k^2} \int_{-1}^{1} \| \frac{\partial }{\partial y} \hat{v} \|^2 + k^2 \| \hat{v} \|^2 + \| \hat{\eta} \|^2 dy\ d\alpha \ d \beta \text{.}
\end{equation}
\begin{figure}
    \centering
    \begin{overpic}{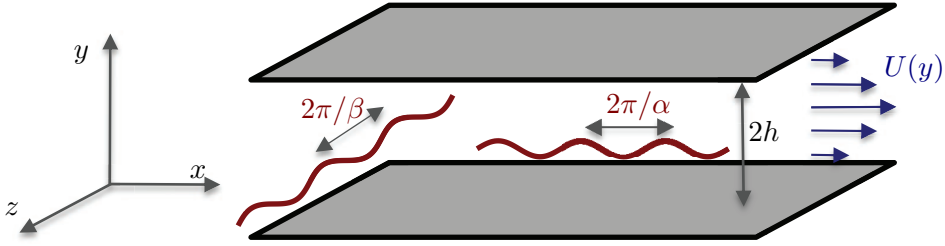}
    \put(20,12.5){ $x$}
    \put(7.5,25.5){ $y$}
    \put(0,8.5){ $z$}
    \put(80.3,16.5){ $2h$}
    \put(66,19.5){\color{myred} $2\pi/\alpha$}
    \put(33,19.0){\color{myred} $2\pi/\beta$}
    \put(96,23.5){\color{myblue} $U(y)$}
    \end{overpic}
    \caption{Schematic of Poisseuille flow. The red waves represent disturbances with particular wavenumbers.}
    \label{fig:my_label}
\end{figure}
\\

\section{Optimal transient growth} \label{sec:Transient_Growth}
Here, we review the linear effects responsible for transient growth in a system with all negative eigenvalues. For a more thorough review, see Ref. \cite{Schmid07}. Expressing the Navier-Stokes equations as
\begin{equation}
    \dot{\boldsymbol{q}}(\boldsymbol{x},t) =  \mathcal{N}(\boldsymbol{q}(\boldsymbol{x},t)) \text{,}
\end{equation}
a steady solution $\overline{\boldsymbol{q}}(\boldsymbol{x})$ satisfies $\mathcal{N}(\overline{\boldsymbol{q}}(x)) = \boldsymbol{0}$. Though their size is likely small, disturbances to the base flow are inevitable. Denoting these disturbances as $\boldsymbol{q}(\boldsymbol{x},t)$, their dynamics are analyzed by linearizing around the base flow,
\begin{equation} \label{eq:TG:lin_approx}
     \frac{d \boldsymbol{q}(\boldsymbol{x},t)}{dt} = \mathcal{N}(\overline{\boldsymbol{q}}(x) + \boldsymbol{q}(\boldsymbol{x},t)) \approx \boldsymbol{A}\boldsymbol{q}(\boldsymbol{x},t) \text{,}
\end{equation}
where $\boldsymbol{A}$ is the Jacobian around the base flow,
\begin{equation}
    \boldsymbol{A} = \frac{\partial \mathcal{N}}{\partial \boldsymbol{q}} \big|_{\overline{\boldsymbol{q}} } \text{.}
\end{equation}
The problem is discretized as 
\begin{equation} \label{eq:TG:ODE}
    \dot{\bf{q}}(t) = {\bf{A}}{\bf{q}}(t) \text{,}
\end{equation}
where ${\bf q}(t) \in \mathbb{R}^N$ is the discretized state vector describing the disturbance. 
\\

The solution to (\ref{eq:ODE_sol}) is
\begin{equation} \label{eq:ODE_sol}
    {\bf{q}}(t) = {\bf M}_t{\bf{q}}(0) \text{,}
\end{equation}
where the evolution operator is the matrix exponential
\begin{equation}
    {\bf M}_t = \exp[{\bf{A}}t] \text{.}
\end{equation}
If all of the eigenvalues of the linear operator $\bf A$ have a negative real part, then the linear system is stable in the sense that the norm of any disturbance will eventually decay, i.e., $\lim_{t\to \infty} \|{\bf q}(t)\| = 0$. This sense of stability, usually referred to as modal stability, is mathematically powerful --- it is a property of the system, not of any particular disturbance. If the eigenvalues are negative, any disturbance decays eventually, but if there is a positive eigenvalue, any disturbance arising in a physical scenario will have a non-zero projection onto the associated eigenvector, and will thus grow exponentially. 
\\

The theory of transient growth offers the additional insight that even if all the eigenvalues are stable, if ${\bf A}$ is non-normal, i.e., its eigenvectors are non-orthognoal, the decay need not be monotonic. The eigenvectors summed together to construct an initial disturbance may mostly cancel each other initially, but because they vanish at different rates, after some time, there may no longer be cancellation, which leads to growth of the disturbance. The linear operators arising in fluids systems, especially in shear flows, can be highly non-normal \cite{Trefethen91}. The ability for these systems to produce growth is quantified in the literature by the maximal amplification that a disturbance may undergo,
\begin{equation} \label{eq:TG:gopt}
    G^{opt}(t) \equiv \max_{\|{\bf q}(0)\| = 1} \|{\bf q}(t)\|^2 \text{.}
\end{equation}
\\

This quantity is usually referred to simply as $G$. Here, we have termed it $G^{opt}$ to specify that it is the optimal growth among all possible initial disturbances and to distinguish it from $G^{mean}$, which will arise later in the paper. Its peak in time is referred to in this paper as $G^{opt}_{max}$ (usually referred to simply as $G_{max}$). The norm $\| \cdot \|$ is based on the kinetic energy of the disturbance and can be written
\begin{equation} \label{eq:TG:energy_weight_def}
    e({\bf q}) = \|{\bf q} \|^2 = {\bf q}^* {\bf W q} \text{.}
\end{equation}
${\bf W}$ is a weight matrix (required to be Hermetian and positive-definite), and we make frequent use of the decomposition ${\bf L}^* {\bf L} = {\bf W}$. For later use, the inner product that induces the norm is $\langle {\bf q}_1 , {\bf q}_2 \rangle = {\bf q}_2^* {\bf W} {\bf q}_1$. It can be shown that the optimal growth may be written \cite{Reddy93}
\begin{equation}
    G^{opt}(t) = \sigma_1^2({\bf L M}_t {\bf L}^{-1}) \text{,}
\end{equation}
where $\sigma_1^2(\cdot)$ returns the first (squared) singular value of the argument. The structures that undergo the most growth up to time $t$ and the structures resulting from the amplification may also be obtained via the singular value decomposition of the weighted evolution operator,
\begin{equation} \label{TG:SVD}
    {\bf L M}_t {\bf L}^{-1} = \tilde{\bf U}{\bf \Sigma}\tilde{\bf V}^{*}\text{.}
\end{equation}
The optimal output and input modes are recovered as ${\bf U} = {\bf L}^{-1}\tilde{\bf U}$ and ${\bf V} = {\bf L}^{-1}\tilde{\bf V}$, respectively. The first column of ${\bf V}$ is the initial disturbance that grows by $G^{opt}(t)$, and the first column of ${\bf U}$ is the structure that results.
\\

The largest initial growth rate experienced by any disturbance can be expressed in terms of the optimal growth as 
\begin{equation}
a^{opt} = \frac{d}{dt} G^{opt}(t) \Big|_{t = 0} \text{.}
\end{equation} 
By expanding the matrix exponential to first order terms in $t$, it is easily shown that this optimal growth rate is given by the numerical abscissa \cite{Trefethen05}
\begin{equation}
a^{opt} = \kappa_1 ( {\bf L A L}^{-1} + ({\bf L A L}^{-1})^* ) \text{,}
\end{equation}
where $\kappa_1(\cdot)$ returns the first eigenvalue of the argument.
\\

\begin{figure} [!hbt]
    \centering
    \input{Figures_2/TG_gopt.tex}
    \includegraphics{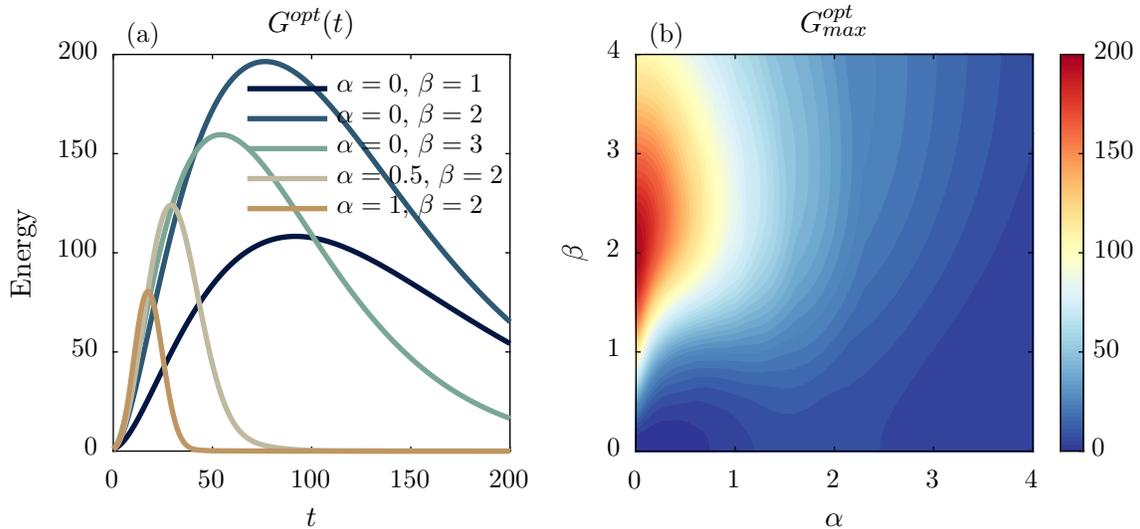}
    \caption{Optimal gains for plane Poisseuille flow at $Re = 1000$: (a) The maximal gain over all initial disturbances $G^{opt}(t)$ for various choices of wavenumbers. (b) Maximal gain, also maximized over time for a range of streamwise and spanwise wavenumbers $\alpha$, $\beta$.}
    \label{fig:TG:Gmax}
\end{figure}
So long as the disturbance remains small enough, the linear approximation (\ref{eq:TG:lin_approx}) remains valid, and the disturbance will decay to zero. However, if the growth is large enough, it can elevate a disturbance from the regime where linearity governs to one where nonlinear effects are relevant. These nonlinear effects can in turn lead the flow away from the base state, eventually causing transition. The growth can indeed be quite large, owing to the severe non-normality in the linearized Navier-Stokes operator in shear flows. Figure \ref{fig:TG:Gmax}(a) shows $G^{opt}(t)$ for various streamwise and spanwise wavenumbers in plane Poisseuille flow at $Re = 1000$. For $\alpha = 0$, $\beta = 2$, $G_{max}^{opt}$ is nearly $200$. Figure \ref{fig:TG:Gmax}(b) shows $G_{max}^{opt}$ for a range of wavenumbers. Streamwise-elongated structures (small $\alpha$) are capable of larger growth than shorter structures (larger $\alpha$). The peak in wavenumber space is at $\alpha = 0$, $\beta = 2.04$, so structures of finite spanwise ($z$) length experience the most growth.
\\

\begin{figure}
    \centering
    \input{Figures_2/trajs_nolevels.tex}
    \includegraphics{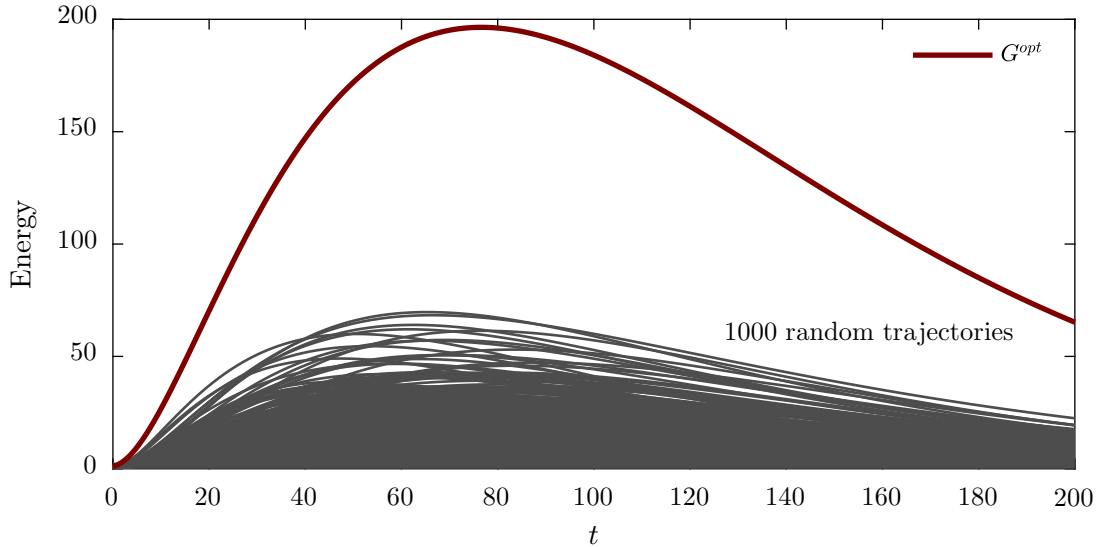}
    \caption{$G^{opt}(t)$ for $\alpha = 0$, $\beta = 2$ along with $1000$ random trajectories.}
    \label{fig:TG:trajs}
\end{figure}
To motivate the remainder of this paper, we show $1000$ random trajectories along with $G^{opt}(t)$ at $Re = 1000$, $\alpha = 0$, $\beta = 2$ in Figure \ref{fig:TG:trajs}. $G^{opt}$ indeed bounds the trajectories, but, notably, they all substantially undershoot it. The details of the distribution used to generate Figure~\ref{fig:TG:trajs} are given in \S~\ref{sec:Estimate_PDF}. In what follows, we derive formulae to describe the statistics of the growth and demonstrate them on plane-Poisseuille flow, recording our observations.
\\

\section{Expected energy amplification} \label{sec:exp_amplification}
In light of Figure~\ref{fig:TG:trajs}, an obvious question is: how much energy, on average, do the amplified disturbances achieve? We derive a formula for the mean energy of the amplified disturbances in terms of the correlation matrix of the initial disturbances. The expected energy divided by the expected initial energy is termed $G^{mean}$. We elaborate on the difference between this and the expected value of the ratio of these energies at the end of the following subsection, but, in short, $G^{mean}$ is more physically meaningful, produces a simpler mathematical result, and requires less a priori knowledge of the initial disturbances.
\\

Just as in the standard treatment of transient growth, the physical model that we consider consists of the discretized base flow $\overline{\bf q}$, which is impulsively perturbed at $t=0$ by ${\bf q}(0)$. As before, the disturbance ${\bf q}(0)$ may represent the entire (three-dimensional) flow in space, the flow at a particular pair of wavenumbers, or the flow at a particular location in the streamwise direction. The evolution of the disturbances is governed by the Navier-Stokes equations linearized around the base flow, so (\ref{eq:ODE_sol}) holds. However, our statistical framework differs from the standard treatment of transient growth in that the disturbance ${\bf q}(0)$ is now a random variable with some distribution, and we study the statistics of the disturbance after some time ${\bf q}(t)$. In particular, we are interested in the energy of the growing disturbance in comparison to that of the initial disturbance.
\\

Experimenting with various choices of the initial correlation for plane-Poisseuille reveals that the expected energy can be substantially smaller than $G^{opt}_{max}$. This is especially true when the correlation length is short relative to the channel half-height. Furthermore, $G^{mean}_{max}$ drops off more rapidly with larger $\alpha$ than does $G^{opt}_{max}$, which causes the mean energy amplification for three-dimensional disturbances to be quite small unless their energy is focused sharply at $\alpha = 0$. Surprisingly, we observe that for isotropically correlated initial disturbances, the mean energy amplification scales near-linearly with $Re$, in contrast to the quadratic scaling of $G^{opt}_{max}$.
\\

\subsection{Theory}
\subsubsection{$G^{mean}$}
For simplicity, we omit the weight matrix in the derivations (by setting it to the identity), reporting the formulae with it at the end, so $e({\bf q}(t)) = {\bf q}^*(t){\bf q}(t)$. The energy may alternatively be written as the trace of the outer product,
\begin{equation} \label{eq:gmean:energy_trace}
    e({\bf q}(t)) = \Tr \{ {\bf q}(t) {\bf q}^*(t) \} \text{,}
\end{equation}
because the diagonals of ${\bf q}(t) {\bf q}^*(t)$ are the terms summed in the inner product. In terms of the evolution operator, (\ref{eq:gmean:energy_trace}) becomes
\begin{equation}
    e({\bf q}(t)) = \Tr \{ {\bf M}_t{\bf q}(0) {\bf q}^*(0){\bf M}^*_t \} \text{.}
\end{equation}
The expected value of this expression gives the expected energy of the amplified disturbances,
\begin{equation}
    \mathbb{E}[e({\bf q}(t))] = \mathbb{E} [ \Tr \{ {\bf M}_t{\bf q}(0) {\bf q}^*(0){\bf M}^*_t \} ] \text{.}
\end{equation}
The expectation commutes with the trace and evolution matrices, giving
\begin{equation}
    \mathbb{E}[e({\bf q}(t))] = \Tr \{ {\bf M}_t\mathbb{E}[{\bf q}(0) {\bf q}^*(0)]{\bf M}^*_t \}  \text{.}
\end{equation}
The expectation of the outer product of the initial disturbances is their correlation matrix,
\begin{equation}
    {\bf C}_{00} = \mathbb{E}[{\bf q}(0) {\bf q}^*(0)] \text{,}
\end{equation}
so the expected energy of the growing disturbance is expressed in terms of the correlations of the initial disturbances,
\begin{equation} \label{exp_energy_1}
    \mathbb{E}[e({\bf q}(t))] = \Tr \{ {\bf M}_t{\bf C}_{00}{\bf M}^*_t \} \text{.}
\end{equation}
A metric for the expected growth of the disturbances, which we term $G^{mean}$, is provided by the ratio of the expected energy and initial energy,
\begin{equation} \label{Gmean}
    G^{mean}(t) \equiv \frac{ \mathbb{E}[e({\bf q}(t))]}{\mathbb{E}[e({\bf q}(0))]} = \frac{\Tr \{ {\bf M}_t{\bf C}_{00}{\bf M}_t^* \}}{ \Tr \{{\bf C}_{00} \}} \text{.}
\end{equation}
This quantity is not the same as the expected value of the growth; this difference is discussed at the end of this subsection. If a weight matrix ${\bf W} = {\bf L}^*{\bf L}$ is used to define the energy, then (\ref{Gmean}) becomes
\begin{equation} \label{Gmean_weight}
    G^{mean}(t) = \frac{\Tr \{ {\bf L} {\bf M}_t {\bf C}_{00}{\bf M}^*_t {\bf L}^* \} }{\Tr \{ {\bf L}^* {\bf C}_{00}{\bf L} \} } \text{.}
\end{equation}
\\

$G^{opt}(t)$ is given in terms of the SVD of the evolution operator. To express $G^{mean}(t)$ in a similar manner, we make use of the fact that the trace of a matrix is the sum of its eigenvalues and that the eigenvalues of ${\bf BB}^*$ are the squared-singular values of ${\bf B}$ for any matrix ${\bf B}$. Using these two facts, (\ref{Gmean}) can be written
\begin{equation} \label{eq:meancalc:svd_form}
    G^{mean}(t) = \frac{\sum_{i=1}^{N}\sigma^2_i({\bf M}_t{\bf B})}{\sum_{i=1}^{N}\sigma_i^2({\bf B})} \text{,}
\end{equation}
where ${\bf B}$ is defined by the factorization ${\bf C}_{00} = {\bf B B}^*$. In the case of a weight matrix, (\ref{eq:meancalc:svd_form}) becomes
\begin{equation}
    G^{mean}(t) = \frac{\sum_{i=1}^{N}\sigma_i^2({\bf L} {\bf M}_t{\bf B})}{\sum_{i=1}^{N}\sigma_i^2({\bf L}{\bf B})} \text{.}
\end{equation}
Upper and lower bounds for $G^{mean}(t)$ for any possible initial correlation can be obtained by setting ${\bf C}_{00}$ to the outer product of the first input mode with itself and last input mode with itself, i.e., ${\bf v}^1 {\bf v}^{1*}$ and ${\bf v}^N {\bf v}^{N*}$, respectively, yielding the bounds $\sigma_1^2({\bf LM}_t {\bf L}^{-1})$ and $\sigma_N^2({\bf LM}_t {\bf L}^{-1})$. Notably, the upper bound is $G^{opt}(t)$. In the case that the disturbances are white in space, i.e, ${\bf C}_{00} = {\bf W}^{-1}$, the resulting $G^{mean}(t)$ is the mean-squared singular value of the weighted evolution operator ${\bf L M}_t{\bf L}^{-1}$.
\\

$G^{mean}$, defined in (\ref{Gmean}), is the ratio of the expected energy of the disturbance at time $t$ to its expected initial energy. This is distinct from the expected ratio of energy, $\mathbb{E}\big[\frac{e({\bf q}(t))}{e({\bf q}(0))}\big]$. Physically, the ratio of expected energies is the more salient quantity because whether a particular disturbance leads to transition depends on its final energy (and shape), not on the growth it underwent. In Figure \ref{fig:TG:trajs}, this ratio of expected energies is the mean of the gray curves (at each time). The expected ratio of energies would come from dividing each disturbance by its initial energy, then taking the average, but this inappropriately weights the growth of smaller initial disturbances equal to that of larger ones. Mathematically, the ratio of expected energies is the easier quantity to work with because it depends only on the correlation matrix of the initial disturbances, as shown in (\ref{Gmean}), while the expected ratio of energies depends on the entire distribution of the initial disturbances. If there is no variation in the size of initial disturbances, i.e., if they live on an $N$-dimensional sphere, the two quantities are the same. More generally, the quantities are the same in the case that the distribution of initial disturbances is separable in radius and direction, as is proven in Appendix~\ref{App:first_appendix}.
\\

Analogous to the numerical abscissa $a^{opt}$, we define the mean initial growth rate,
\begin{equation}
    a^{mean} \equiv \frac{d}{dt}G^{mean}(t) \Big|_{t = 0} \text{.}
\end{equation}
This derivative can be calculated by expanding the evolution operator to first order,
\begin{equation}
a^{mean} = \frac{d}{dt} \frac{ \Tr \left\{ {\bf L} ( {\bf I} + t{\bf A}) {\bf C}_{00}({\bf I} + t{\bf A}^* ){\bf L}^* \right\} } {\Tr \{ {\bf L} {\bf C}_{00} {\bf L}^* \} } \bigg|_{t = 0}\text{,}
\end{equation}
where ${\bf I}$ is the identity. Dropping the quadratic term and evaluating the derivative gives
\begin{equation}
    a^{mean} = \frac { \Tr \{ {\bf L}{\bf A C}_{00}{\bf L}^* + {\bf LC}_{00}{\bf A}^* {\bf L}^* \} } {\Tr \{ {\bf L} {\bf C}_{00} {\bf L}^* \} } \text{.}
\end{equation}
Finally, leveraging the Hermicity of the correlation matrix,
\begin{equation} \label{eq:amean}
    a^{mean} = 2\frac {  \Tr \{ \mathrm{Re} ({\bf L}{\bf A C}_{00}{\bf L}^* )\}} {\Tr \{ {\bf L} {\bf C}_{00} {\bf L}^* \} } \text{,}
\end{equation}
where $\mathrm{Re}(\cdot )$ returns the real part of the argument. The upper bound for this quantity is $a^{opt}$, which is positive if (and only if) $G^{opt}>1$, but we have never observed  $a^{opt}$ to be positive in our numerical experiments. Indeed, we have never observed a randomly chosen disturbance initially grow. 
\\

\subsubsection{Correlation and dominant structures} \label{Theory:POD}
The statistics of the initial disturbances can also be used to augment prediction of the structures that arise from the linear amplification by the evolution operator. Removing the trace from (\ref{exp_energy_1}) gives a formula for the correlation matrix of the disturbance at time $t$,
\begin{equation} \label{Correlation_out}
     {\bf C}_{tt} \equiv \mathbb{E}[{\bf q}(t){\bf q}(t)^*] = {\bf M}_t{\bf C}_{00}{\bf M}_t^* \text{.}
\end{equation}
The dominant flow structures at time $t$ are the eigenvectors of this correlation matrix (multiplied by a weight if desired),
\begin{equation} \label{POD:mode_def}
    {\bf C}_{tt} {\bf W}{\bf \Phi}_t = {\bf \Phi}_t{\bf \Lambda}_t \text{.}
\end{equation}
The columns $\boldsymbol{\phi}^k_t$ of ${\bf \Phi}_t$ are orthogonal in the weighted inner product, i.e., $\langle \boldsymbol{\phi}^i_t,\boldsymbol{ \phi}^j_t \rangle = \delta_{ij}$. This can be thought of as a particular variant of proper orthogonal decomposition (POD)   \cite{Lumley70, Lumley67,Sirovich87} in which the data consists of an ensemble of realizations of the disturbances at a specific time $t$ rather than a single time series. The eigenvalues are non-negative, owing to the semi-positive definiteness of the correlation matrix, and represent the expected energy of each structure. More precisely, the $k$-th eigenvalue 
\begin{equation}
    \lambda^k_t = \mathbb{E}[| \langle {\bf q}(t) , \boldsymbol{\phi}^k_t \rangle|^2] \text{}
\end{equation}
is the expected energy of the projection of the disturbance onto the $k$-th mode $\boldsymbol{\phi}^k_t$. The eigenvalues sum to the total expected energy, so
\begin{equation} \label{eq:thry:Gmean_eigenvalues}
   G^{mean}(t) = \frac{\sum_i \lambda_t^i }{\sum_i \lambda_0^i} \text{.}
\end{equation}
Therefore, the eigenvalues quantify the expected contribution of each mode to the growth of the disturbance. 
\\

The average energy of the disturbance captured by any structure can be quantified \cite{Frame22} by,
\begin{equation} \label{eq:theory:POD:energy_capture}
    \epsilon(\boldsymbol{\psi}) = \mathbb{E}[| \langle  {\bf q}(t) , \boldsymbol{\psi} \rangle |^2] \text{.}
\end{equation}
The first POD mode maximizes this quantity (over normalized modes), and the latter modes maximize it with the constraint that they are orthogonal to all previous ones. For a more thorough review of POD, see Refs. \cite{Rowley17,Taira18,Towne18}. 
\\

The POD modes offer an alternative to the output modes of the evolution matrix for describing the structures that emerge from the linear amplification. The POD modes are the most energetic structures, while the output modes are the modes resulting from the greatest amplification by the evolution operator. In the case that the initial correlation ${\bf C}_{00}$ is white with respect to the weight, the POD modes are equivalent to the output modes, i.e., 
\begin{equation}
    {\bf C}_{00} = {\bf W}^{-1} \implies {\bf \Phi} = {\bf U} \text{.}    
\end{equation}
This result is analogous to the relationship between resolvent modes and spectral POD modes established in Ref. \cite{Towne18}. Of course, the initial correlation is unlikely to be white in a real flow, so it is advantageous to use knowledge of the incoming statistics to augment the prediction of these structures. 
\\

In the remainder of this section, we experiment with different choices of ${\bf C}_{00}$ for Poisseuille flow and record our observations. We are not aware of any previous studies on the correlations of disturbances within Poisseuille flow. The nature of the disturbances and quantities such as their correlations are certainly sensitive to the specifics of the flow setup. For example, the disturbances generated by vibrations of the boundary are likely substantially different from those caused by surface roughness. Providing a model for the correlations of the initial disturbance is not the topic of this paper, and we do not claim that the choices made are necessarily reflective of the physics in Poisseuille flow. However, trends that emerge, e.g., that longer correlation lengths lead to more growth and that $G^{mean}$ is substantially smaller than $G^{opt}$, are not specific to our choice of the correlation, and therefore give physical insight despite the current lack of an accurate model for the correlations.

\subsection{Numerical experiments with disturbances at a single wavenumber pair} \label{subsec:single_ab}
Most studies of transient growth in flows with homogeneous directions take the Fourier transform in these directions and calculate the transient growth for disturbances consisting of a single pair of streamwise and spanwise wavenumbers. Here we perform the analogous analysis for $G^{mean}$ in Poisseuille flow. The correlation at a particular $\alpha$ and $\beta$ can be written
\begin{equation}
    \hat{\boldsymbol{C}}_{00}(y_1,y_2;\alpha,\beta) = \begin{bmatrix} \hat{\boldsymbol{C}}_{00}^{vv}(y_1,y_2;\alpha,\beta) && \hat{\boldsymbol{C}}_{00}^{v\eta}(y_1,y_2;\alpha,\beta) \\ \hat{\boldsymbol{C}}_{00}^{\eta v}(y_1,y_2;\alpha,\beta) && \hat{\boldsymbol{C}}_{00}^{\eta \eta}(y_1,y_2;\alpha,\beta)
    \end{bmatrix} \text{,}
\end{equation}
where the diagonal terms are the autocorrelations of wall-normal velocity and wall-normal vorticity, and the off diagonal terms are the cross correlations between these two variables. It can be shown analytically that for a disturbance to experience large growth, its initial energy should be concentrated in its wall-normal velocity rather than wall-normal vorticity \cite{Gustavsson91}. Therefore, we choose only the vertical velocity autocorrelation to be non-zero and take it to be Gaussian in the wall-normal direction with correlation length $\lambda$, i.e., 
\begin{equation} \label{eq:singleab:correlation_form}
    \hat{\boldsymbol{C}}_{00}^{vv}(y_1,y_2;\alpha,\beta) =  \mathbb{E} [\hat{v}(y_1,\alpha,\beta)\hat{v}(y_2,\alpha,\beta)] = A\exp \left[-\frac{(y_1-y_2)^2}{\lambda^2} \right] \text{.}
\end{equation}
The normalization $A$ has no impact on $G^{mean}$ because this constant affects the expected energy of the amplified disturbances and that of the initial ones equally. In our numerics, it is chosen so that when the initial correlation is discretized in $y$, its trace is unity.
\\

\subsubsection{$G^{mean}$ for a single wavenumber pair}
\begin{figure} [!hbt]
    \centering
    \input{Figures_2/gmean_time.tex}
    \includegraphics{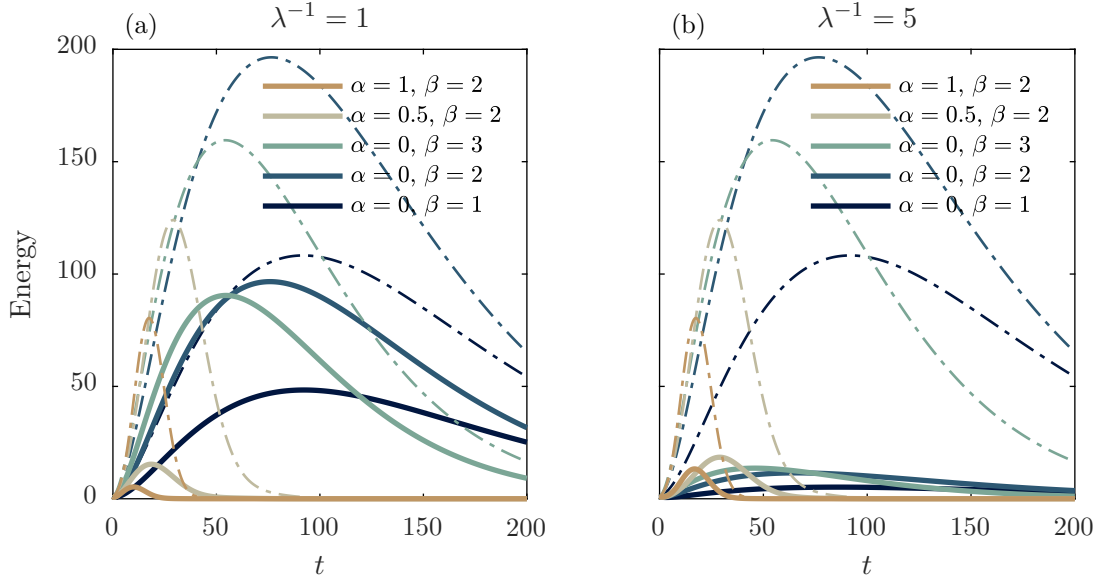}
    \caption{$G^{mean}(t)$ (solid) and $G^{opt}(t)$ (dashed) for various wavenumber  at $Re = 1000$. The mean energy amplification is substantially higher for the longer correlation length $\lambda^{-1} =  1$ than for the shorter one $\lambda^{-1} =  5$.}
    \label{fig:single_ab:gmean_time}
\end{figure}

Figure \ref{fig:single_ab:gmean_time} shows $G^{mean}(t)$ (solid) for various wavenumbers and $G^{opt}(t)$ (dashed) for the same wavenumbers, both as functions of time for $Re = 1000$. Whether the mean is on the same order as the maximum depends on the characteristics of the correlations of the initial disturbances. We refer to the peak of $G^{mean}(t)$ in time as $G^{mean}_{max}$. For the relatively long correlations in (a), $G^{mean}_{max}$ is roughly half $G^{opt}_{max}$ for the most amplified wavenumbers, while for the shorter correlation length (b), the ratio is closer to one-tenth. 
\begin{figure}[!h]
    \centering
    \input{Figures_2/gmean_time_zoom.tex}
    \includegraphics{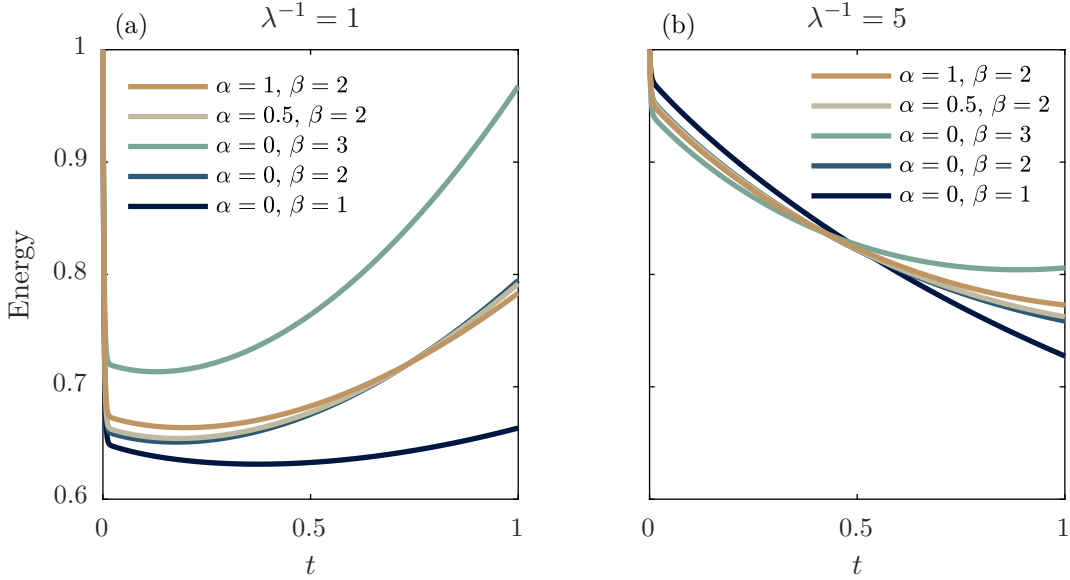}
    \caption{$G^{mean}(t)$ for short times. A steep decay is observed initially even in cases where $G^{mean}_{max}$ is relatively high. The initial growth (or decay) rate is $a^{mean}$,  given in (\ref{eq:amean}).}
    \label{fig:single_ab:gmean_time_zoom}
\end{figure}
Figure~\ref{fig:single_ab:gmean_time_zoom} shows the first time unit of $G^{mean}(t)$ using the same parameters as Figure~\ref{fig:single_ab:gmean_time}. Despite the fact that $G^{mean}$ grows to be relatively large, it initially decays sharply for all wavenumbers. The initial decay rate can be calculated with (\ref{eq:amean}). 
\begin{figure}[!h]
    \centering
    \input{Figures_2/gmean_max_vslam.tex}
    \includegraphics{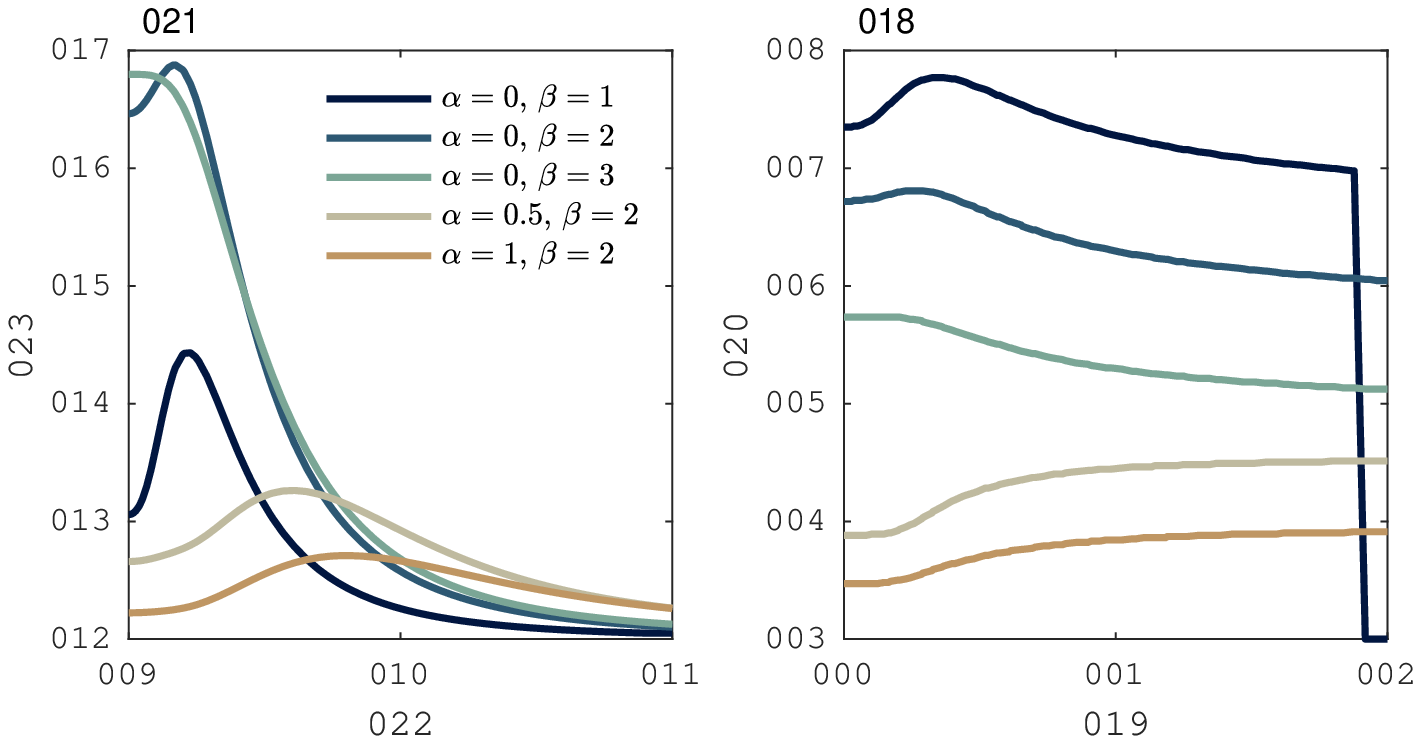}
    \caption{The effect of correlation length on $G^{mean}_{max}$ for Poisseuille flow at $Re = 1000$.  (a) $G^{mean}(t)$ maximized over time vs. inverse correlation length for various streamwise and spanwise wavenumbers. More coherent disturbances (large $\lambda$) tend to grow more, but there is a non-infinite optimum. (b) The time at which $G^{mean}$ is maximized vs. inverse correlation length. The maximum time does not vary much with $\lambda$ but does with $\alpha$ and $\beta$, with shorter wavenumbers corresponding to an earlier maximization time. The maximization time drops to zero when the correlation becomes such that $G^{mean}(t)$ never exceeds $1$.}
    \label{fig:single_ab:lam_sweep}
\end{figure}
\\

Figure \ref{fig:single_ab:lam_sweep}(a) shows $G^{mean}_{max}$ for a range of $\lambda^{-1}$ at $Re = 1000$. The correlation length $\lambda$ greatly impacts the mean energy amplification, with longer correlation lengths corresponding to more growth and shorter ones to less growth. It is likely this trend is explained by the fact that short-wavelength (in $y$) disturbances are quickly dissipated by viscosity before they can extract energy from the mean shear \cite{McKeon17}. Figure \ref{fig:single_ab:lam_sweep}(b) shows the time at which $G^{mean}$ is maximized. This time is relatively independent of the correlation length, but changes substantially with the wavenumber pair.
\\

\begin{figure} [!hbt]
    \centering
    \input{Figures_2/Contour_ab_multilam.tex}
    \includegraphics{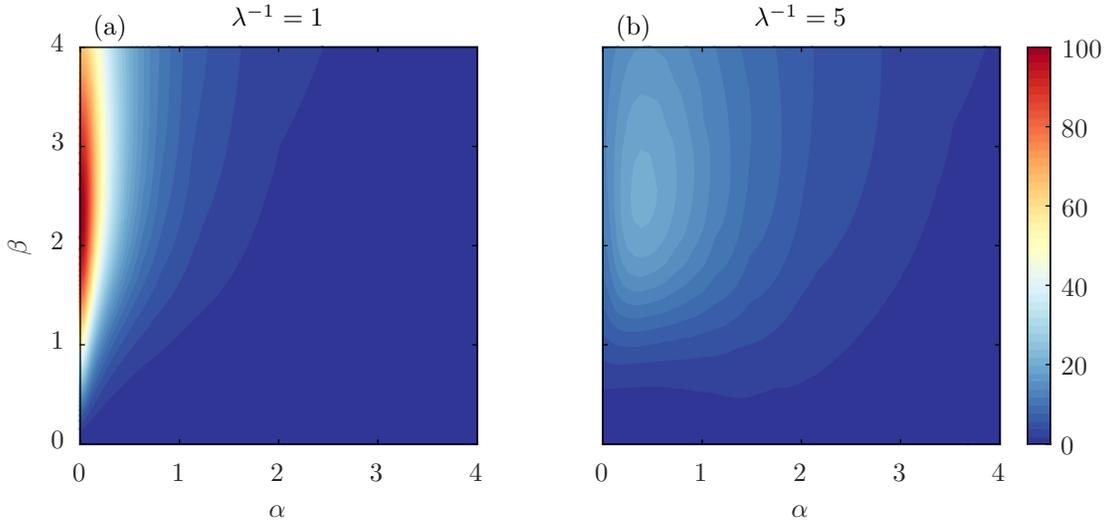}
    \caption{Dependence of $G^{mean}$ on the streamwise and spanwise wavenumbers at $Re = 1000$ for different correlation lengths. The shape is similar to the contour of $G^{opt}_{max}$ at the same Reynolds number (Figure \ref{fig:TG:Gmax}), but, notably, the support in $\alpha$ is substantially narrower for $G^{mean}$. This indicates that the energy of the disturbance must be quite concentrated at the large-growth wavenumbers to achieve significant growth.}
    \label{fig:single_ab:abcontour}
\end{figure}
The wavenumber dependance is further explored in Figure \ref{fig:single_ab:abcontour}. Figure~\ref{fig:single_ab:abcontour} (a) shows this dependence for $\lambda^{-1} = 1$, which is near the peak for the maximally amplified wavenumber in Figure \ref{fig:single_ab:lam_sweep} (a). The location of the peak in wavenumber space is near that of $G^{opt}_{max}$ seen in Figure \ref{fig:TG:Gmax}; however, $G^{mean}_{max}$ decays much more rapidly with $\alpha$ than does $G^{opt}_{max}$. This indicates that to achieve large-scale growth, the energy of a disturbance must be narrowly concentrated in wavenumber-space at the values that experience large growth. As shown in \S~\ref{subsec:multipleab_num_exp}, this severely limits the mean energy amplification of fully three-dimensional disturbances.
\\

\begin{figure}
    \centering
    \input{Figures_2/lamRe_singleab_crit.tex}
    \includegraphics{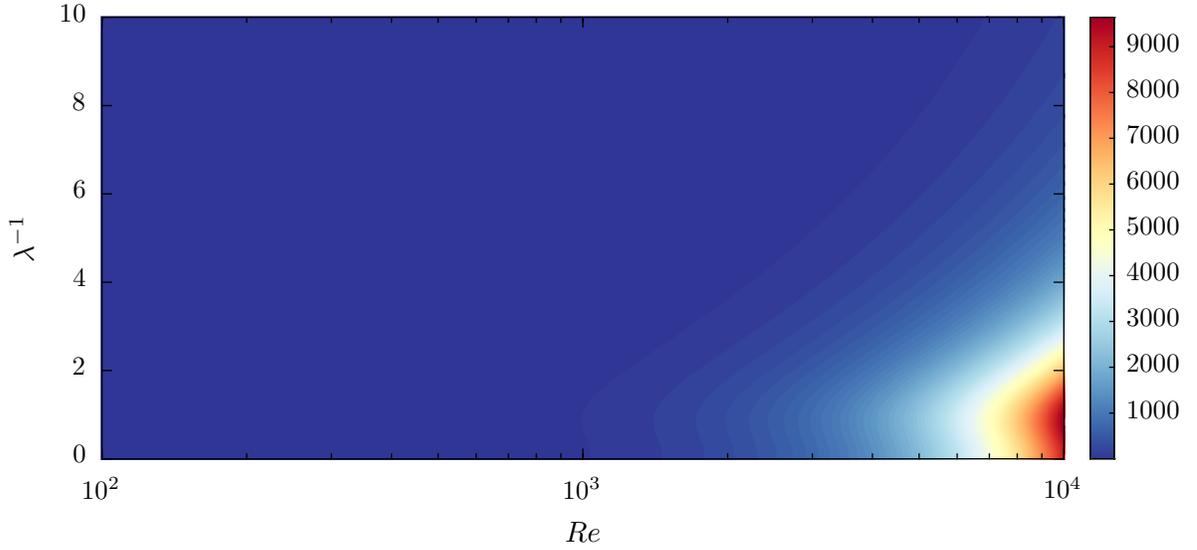}
    \caption{$G^{mean}_{max}$ at $\alpha = 0$, $\beta = 2$ as a function of Reynolds number and inverse correlation length $\lambda^{-1}$. Similar behavior is observed over the range of $Re$.}
    \label{fig:lamRe_sweep_singleab}
\end{figure}
Figure~\ref{fig:lamRe_sweep_singleab} shows $G^{mean}_{opt}$ at $\alpha = 0$, $\beta = 2$ for a range of Reynolds numbers and correlation lengths. Similar dependence on correlation length is observed at all Reynolds numbers with the peak occuring when the correlation length is roughly the channel half-height.
\begin{figure} [!hbt]
    \centering
    \input{Figures_2/1D_Rescaling_crit.tex}
    \includegraphics{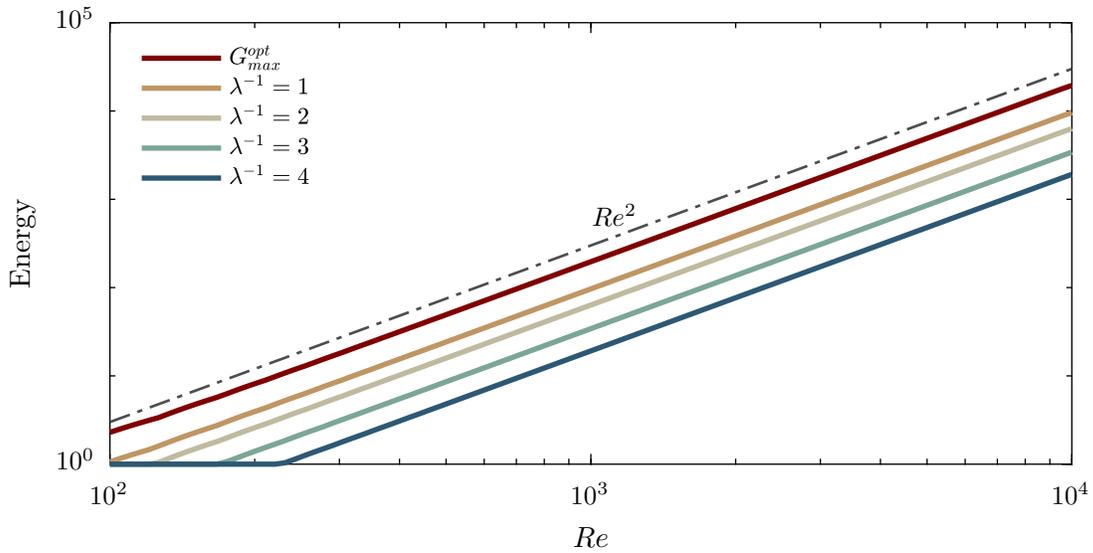}
    \caption{Scaling of $G^{mean}_{max}$ for disturbances at a single wavenumber pair for a variety of correlation lengths at $\alpha = 0$, $\beta = 2$. The scaling appears quadratic (the gray dashed line is $Re^2$), matching that of $G_{max}^{opt}$.}
    \label{fig:singla_ab:re_scaling}
\end{figure}
$G^{opt}_{max}$ is known to scale quadratically with Reynolds number (for small values of $\alpha Re$) \cite{Gustavsson91}. In Figure \ref{fig:singla_ab:re_scaling}, we show the scaling of $G^{mean}_{max}$ with Reynolds number for a variety of correlation lengths at $\alpha = 0$, $\beta = 2$. These appear to obey the same scaling.
\\

\subsubsection{Dominant structures for a single wavenumber pair} \label{subsub:dominant structures}
Now we examine the structures that emerge in Poisseuille flow at a single wavenumber pair, as described in \S~\ref{Theory:POD}. The key question is: to what extent do the output modes resemble the principal components of the correlation matrix, i.e., the POD modes? The former are the structures resulting from the largest amplification by the evolution operator (see (\ref{TG:SVD})), whereas the latter use the statistics of the initial disturbances to inform which structures are most energetic (see (\ref{POD:mode_def})).
\begin{figure} [!hbt]
    \centering
    \input{Figures/correlation_POD.tex}
    \includegraphics{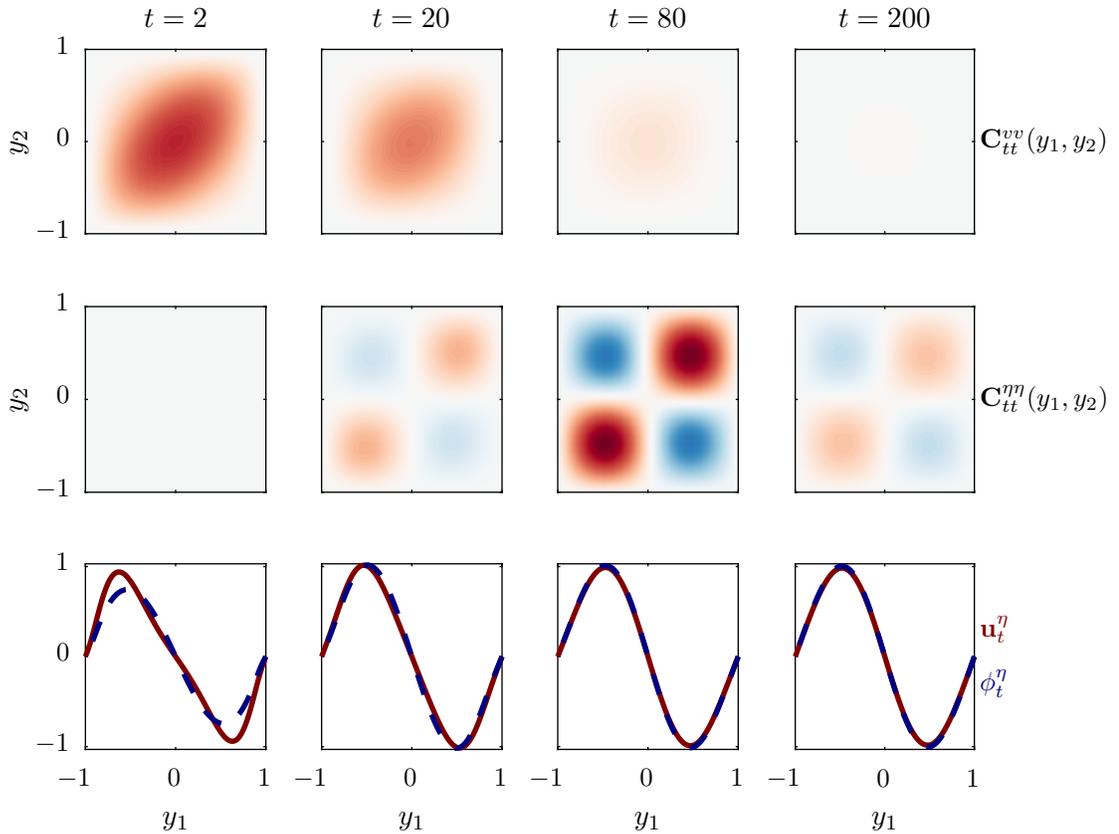}
    \caption{Evolution of correlations and their POD modes for $Re = 1000$, $\alpha = 0$, $\beta = 2$, and $\lambda = 1$. The initial energy is imposed to be in the velocity (top), but it is quickly shifted to vorticity (middle). The first POD mode and first output mode quickly become similar (bottom). For these parameters, $G^{mean}(t)$ peaks near $t = 80$ (see Figure~\ref{fig:single_ab:gmean_time})}.
    \label{fig:singleab:correlation_POD}
\end{figure}
\\

Figure \ref{fig:singleab:correlation_POD} shows the evolution of the correlations and the vorticity component of their POD modes for $\alpha = 0$, $\beta = 2$. We impose the initial correlation to be of the form in (\ref{eq:singleab:correlation_form}) with $\lambda = 1$, so all the energy is initially in the velocity. The evolution operator rapidly shifts this energy to the vorticity, and two counter-rotating vortices emerge. The first POD mode (blue) reflects this with two peaks of opposite sign at the peaks in the vorticity correlation. The velocity component of the POD mode is not plotted because it rapidly decays to $0$. Notably, the first output mode quickly becomes nearly identical to the first POD mode despite the former not depending on the initial correlation. Indeed, for this wavenumber pair, the first few POD modes from different initial correlation matrices quickly become similar to one another and to the first few left singular values of the evolution matrix. There is only moderate gain separation in the singular values, so the similarity between the modes is surprising. 
\begin{figure}
    \centering
    \input{Figures_2/Checkerboard.tex}
    \includegraphics{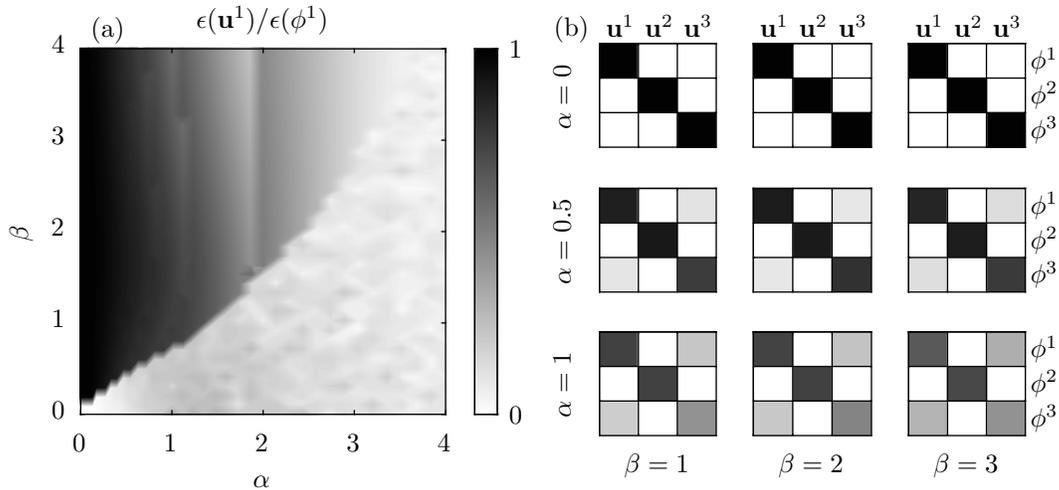}
    \caption{Comparison of the POD and output modes. (a) The ratio of energy captured by the first output mode to that of the first POD mode for a range of wavenumbers. The modes are comparable for low $\alpha$. (b) The square inner products of the first three POD modes and the first three output modes. In both plots, the modes at each wavenumber are compared at the peak time in $G^{mean}$.}
    \label{fig:singleab:checkerboard}
\end{figure}
\\

Figure~\ref{fig:singleab:checkerboard} compares the modes more thoroughly. For each wavenumber pair, Figure~\ref{fig:singleab:checkerboard}(a) shows the average energy captured (see (\ref{eq:theory:POD:energy_capture})) by the first output modes at the peak time of $G^{mean}$ as a fraction of that captured by the first POD mode at the same time. The energy captured by the POD mode is the maximum possible, so a value near unity indicates that the output mode is very effective, while a value near zero indicates the opposite. Whether or not a structure is visible in the flow depends on the energy it captures. Near $\alpha = 0$, the output mode captures nearly as much as is possible, while at higher $\alpha$, it captures substantially less. Capturing energy at low $\alpha$ is more important as the growth is the greatest here, so the leading output mode does a good job of predicting the structures observed in Poisseuille flow \cite{Hiroyuki01}. The large discontinuity in Figure~\ref{fig:singleab:checkerboard}(a) in the lower right of the plot occurs because the peak time for these wavenumbers is $t=0$, so the structures at this time only depend on the initial correlation, not on the linearized Navier-Stokes operator, so the output modes here cannot hope to capture any structure. 
\\

Figure~\ref{fig:singleab:checkerboard}(b) compares the output modes to the POD modes via the inner product, again at the peak time in $G^{mean}$. At each wavenumber pair, the plot shows the matrix of square inner products $G_{ij} = | \langle {\bf u}^i , \boldsymbol{\phi}^j \rangle |^2$ up to three modes in each basis. For $\alpha = 0$, this matrix is nearly diagonal, indicating, once again, that the output modes and POD modes are very similar, even for the subleading modes. At higher $\alpha$, the modes become less similar as shown by the off-diagonal terms in the matrix of inner products.
\\

We also experimented with correlation matrices that do not respect the symmetry about $y = 0$ in the channel (arising, e.g., due to vibrations in one of the plates, but not the other). These resulted in less similarity in the modes, and the square inner products were in the range $0.5-0.9$, even at $\alpha = 0$. However, so long as the correlation was symmetric, the modes were quite similar for $\alpha = 0$. The output modes are therefore a good model for the POD modes under these conditions, and leveraging the statistics may not present much advantage in predicting the structures. We stress, however, that the energy of each structure is highly dependant on the statistics, so the SVD of the evolution operator does not provide the associated energies accurately.  Whether the POD modes and output modes for other flows coincide to the extent that they do in Poisseuille flow may be an interesting topic for future investigation.
\\

It can be shown that the POD modes in flows with homogeneous directions are still delta functions in wavenumber space in those directions \cite{Lumley67}. Therefore, despite the fact that the disturbances will not themselves be delta functions in wavenumber, the similarity in the POD modes and output modes observed in this subsection still applies in the three-dimensional case. However, the behavior of $G^{mean}$ for three-dimensional disturbances can be markedly different, as shown next.
\\

\subsection{Three-dimensional disturbances: inclusion of multiple wavenumbers}
Clearly, just as real initial disturbances will not identically match the maximally amplified one, real disturbances do not exist at just one pair of streamwise and spanwise wavenumbers. Parallel flow offers an analytical simplification to an analysis of transient growth --- each streamwise and spanwise wavenumber pair may be considered separately in its ability to produce growth. However, this tempts further exaggeration of the possibility for large-scale transient growth. In modal stability, one need not add these wavenumbers back together to get an answer as to the long-term behavior --- if any wavenumber pair grows exponentially, so will the entire disturbance. However, if one wavenumber pair experiences large transient growth, this only implies a large gain for the entire disturbance to the extent that its initial energy is concentrated at that wavenumber pair. In this subsection, we incorporate a range of wavenumbers and show that the weight for each pair is determined by the Fourier transform in $x$ and $z$ of the three-dimensional correlation. When these three-dimensional correlations are incorporated, substantially less growth is observed. We also observe a linear scaling with Reynolds number for an isotropic correlation in contrast with the quadratic scaling observed for $G^{opt}_{max}$ and for $G^{mean}_{max}$ at a particular $\alpha$ and $\beta$.  
\\

One way to calculate $G^{mean}$ for disturbances containing multiple wavenumbers would be to define a large domain in $x$ and $z$ (to approximate the desired infinite directions), calculate ${\bf A}$, and define a discrete correlation matrix for the full three-dimensional problem. The mean energy amplification would then be given by (\ref{Gmean}) and the correlation matrix by (\ref{Correlation_out}). However, this strategy is needlessly computationally intensive because it does not take advantage of the analytic simplification possible in parallel flow. Instead, we can add the expected energies at each wavenumber together, modulated by the energy of the incoming disturbances at each wavenumber. Denoting the disturbance discretized in $y$, but continuous in $x$ and $z$ as ${\bf q}_t(x,z)$, its energy is
\begin{align}
\|{\bf q}(t)\|^2 &= \Tr \left\{ \int_{-\infty}^{\infty} {\bf q}_t(x,z){\bf q}_t^*(x,z) dx \ dz \right\}\text{,} \\
 & = \Tr \left\{\int_{-\infty}^{\infty} {\bf q}_t(\alpha,\beta){\bf q}_t^*(\alpha,\beta)  d\alpha \ d\beta \right\} \text{,} \quad \text{(Parseval)} \\
 & = \Tr \left\{\int_{-\infty}^{\infty} {\bf M}_t(\alpha,\beta){\bf q}_0(\alpha,\beta){\bf q}_0^*(\alpha,\beta){\bf M}_t^*(\alpha,\beta)  d\alpha \ d\beta \right\} \text{.}
\end{align}
Finally, taking an expected value, dividing by the expected initial energy, and incorporating the weight matrix yields 
\begin{equation} \label{eq:exenergy:3dgmean}
    G^{mean}(t) = \int_{-\infty}^{\infty} \Tr \left\{{\bf LM}_t(\alpha,\beta)\hat{\bf C}_{00}(\alpha,\beta){\bf M}_t^*(\alpha,\beta){\bf L}^*\right\}  d\alpha \ d\beta  \bigg/ \int_{-\infty}^{\infty} \Tr \left\{ {\bf L} \hat{\bf C}_{00}(\alpha,\beta){\bf L}^*\right\}  d\alpha \ d\beta\text{.}  
\end{equation}
Here, $\hat{\bf C}_{00}(\alpha,\beta) = \mathbb{E}[{\bf q}_0(\alpha,\beta){\bf q}_0(\alpha,\beta)^*]$ is the $y$-discretized correlation of the initial disturbance at each wavenumber pair . By the Wiener-Khinchin theorem \cite{Wiener30,Khintchine34}, $\hat{\bf C}_{00}(\alpha,\beta)$ is, equivalently, the Fourier transform of the three-dimensional correlation, 
\begin{equation}
    \hat{\bf C}_{00}(\alpha,\beta) = \int_{-\infty}^{\infty} {\bf C}_{00}(x,z) e^{-i(\alpha x + \beta z)}dx \ dz \text{.}
\end{equation}
\\

\subsection{Numerical experiments using disturbances with a distribution of wavenumbers} \label{subsec:multipleab_num_exp}
To maximize the potential for growth, we again choose only the wall-normal-velocity autocorrelation to be nonzero.
\subsubsection{Isotropic correlation}
\begin{figure} 
    \centering
    \input{Figures_2/3D_vs_lam.tex}
    \includegraphics{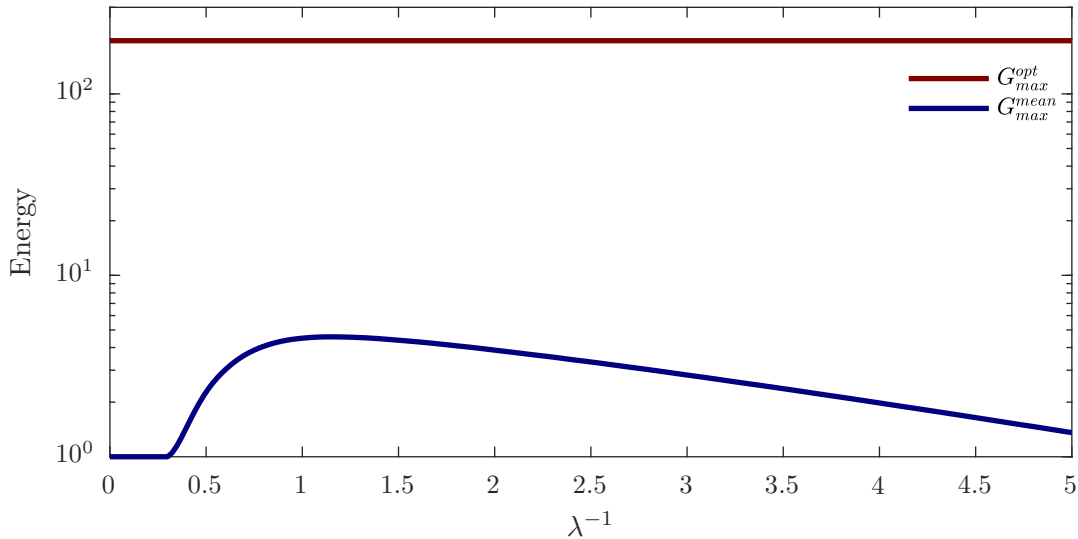}
    \caption{$G^{mean}_{max}$ for a three-dimensional isotropic correlation with correlation length $\lambda$ at $Re = 1000$. Even at the optimal correlation length, the inclusion of all wavenumbers causes $G^{mean}_{max}$ to be roughly $2\%$ of $G^{opt}_{max}$ at $\alpha = 0$, $\beta = 2$.}
    \label{fig:3d_lamsweep}
\end{figure}

We begin by taking the wall-normal autocorrelation to be an isotropic Gaussian with correlation length $\lambda$,
\begin{equation} \label{eq:dist_ab:corr_form}
    \boldsymbol{C}_{00}^{vv}(\boldsymbol{x}_1,\boldsymbol{x}_2) = A\exp \left[-\frac{1}{\lambda^2}(|\boldsymbol{x}_2-\boldsymbol{x}_1|^2) \right] \text{,}
\end{equation}
where $|\cdot|$ denotes Euclidean distance and, again, $A$ has no imact on $G^{mean}$. When discretized in $y$, the correlation becomes ${\bf C}_{00}(\Delta x, \Delta z) = {\bf C}_{00}^{y}\exp[-\frac{1}{\lambda^2}(\Delta x^2 + \Delta z^2)]$, where ${\bf C}_{00}^{y}$ is the discretization of the $y$-dependant part of the correlation with unit trace. The Fourier transform of the correlation in $x$ and $z$ is
\begin{equation} \label{eq:ex:energy:corr_isotropic}
    \hat{\bf C}_{00}(\alpha,\beta) = A \exp \left[-\frac{\lambda^2}{4}(\alpha^2 + \beta^2) \right]{\bf C}_{00}^{y} \text{.}
\end{equation}
Inserting (\ref{eq:ex:energy:corr_isotropic}) into (\ref{eq:exenergy:3dgmean}) gives
\begin{equation}
\begin{aligned}
    G^{mean}(t) = &\int_{-\infty}^{\infty} \Tr \{{\bf LM}_t(\alpha,\beta){\bf C}_{00}^{y}{\bf M}_t^*(\alpha,\beta){\bf L}^*\} \exp\left[-\frac{\lambda^2}{4}(\alpha^2 + \beta^2)\right]  d\alpha \ d\beta \\
    &\bigg/ \int_{-\infty}^{\infty} \Tr \{{\bf LC}_{00}^{y}{\bf L}^*\} \exp \left[-\frac{\lambda^2}{4}(\alpha^2 + \beta^2) \right]  d\alpha \ d\beta \text{.}
\end{aligned}    
\end{equation}
The exponential term can be interpreted as the expected energy at each wavenumber pair implied by the correlation. For Poisseuille flow at $Re = 1000$, the most amplified wavenumbers are near $\alpha = 0,\beta = 2$ (see Figures \ref{fig:TG:Gmax} and \ref{fig:single_ab:abcontour}), and the amplification drops off rapidly as $\alpha$ moves away from zero. To concentrate energy near $\alpha = 0$, the correlation length must be quite long. This longer correlation also promotes growth because, as we described in the previous section, the longer the correlation length in $y$, the more growth is observed. However, with a long correlation length, energy is concentrated near $\beta = 0$, which does not experience much growth (see Figures \ref{fig:TG:Gmax} and \ref{fig:single_ab:abcontour}). Also detracting from $G^{mean}$ is the fact that the maxima occur at significantly different times for different wavenumbers (see Figure~\ref{fig:single_ab:lam_sweep}). The combined effect of these factors can be seen in Figure~\ref{fig:3d_lamsweep}. At $Re = 1000$, even with the correlation length that promotes the most growth ($\lambda^{-1} = 1.2$), $G^{mean}_{max}$ is only $2.5\%$ of $G^{opt}_{max}$ at the optimal wavenumbers. 
\\

\begin{figure} [!hbt]
    \centering
    \input{Figures_2/lamRe_3d_crit.tex}
    \includegraphics{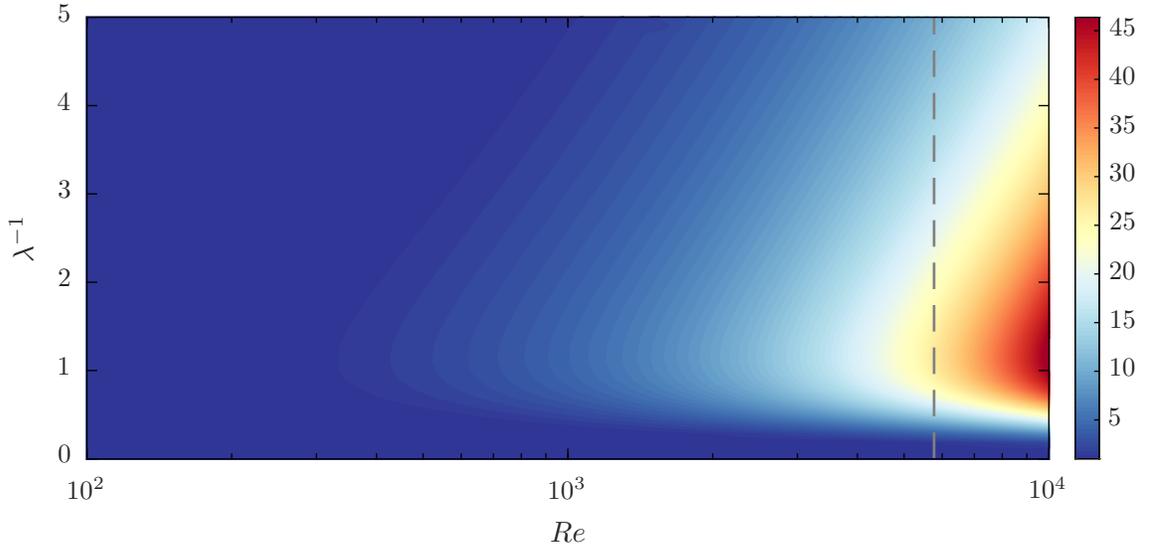}
    \caption{$G_{max}^{mean}$ for an isotropic initial correlation for a range of correlation length and Reynolds number. Similar dependence on correlation length $\lambda$ is observed at all $Re$, and all values are substantially lower than their single-wavenumber counterparts in Figure~\ref{fig:lamRe_sweep_singleab}.}
    \label{fig:multiple_ab:lamRe_contour}
\end{figure}
Figure \ref{fig:multiple_ab:lamRe_contour} shows contours of $G_{max}^{mean}$ for a range of correlation lengths and Reynolds numbers. The vertical dashed line divides the asymptotically stable and unstable regions. At Reynolds numbers higher than this, $G^{mean}_{max}$ is technically infinite, but there is an initial peak in $G^{mean}(t)$ long before the instability dominates. In this figure, and all subsequent ones which show $G^{mean}_{max}$ above the critical Reynolds number, we plot the magnitude of the initial peak in $G^{mean}$. The effect of the correlation length is similar across Reynolds numbers. Comparing Figure~\ref{fig:multiple_ab:lamRe_contour} with Figure~\ref{fig:lamRe_sweep_singleab} (its single-wavenumber analog), we see that the isotropic correlation matrix severely limits growth at all Reynolds numbers. Indeed, the difference becomes greater as the Reynolds number increases.
\\

\begin{figure}
    \centering
    \input{Figures_2/3D_Rescaling_crit.tex}
    \includegraphics{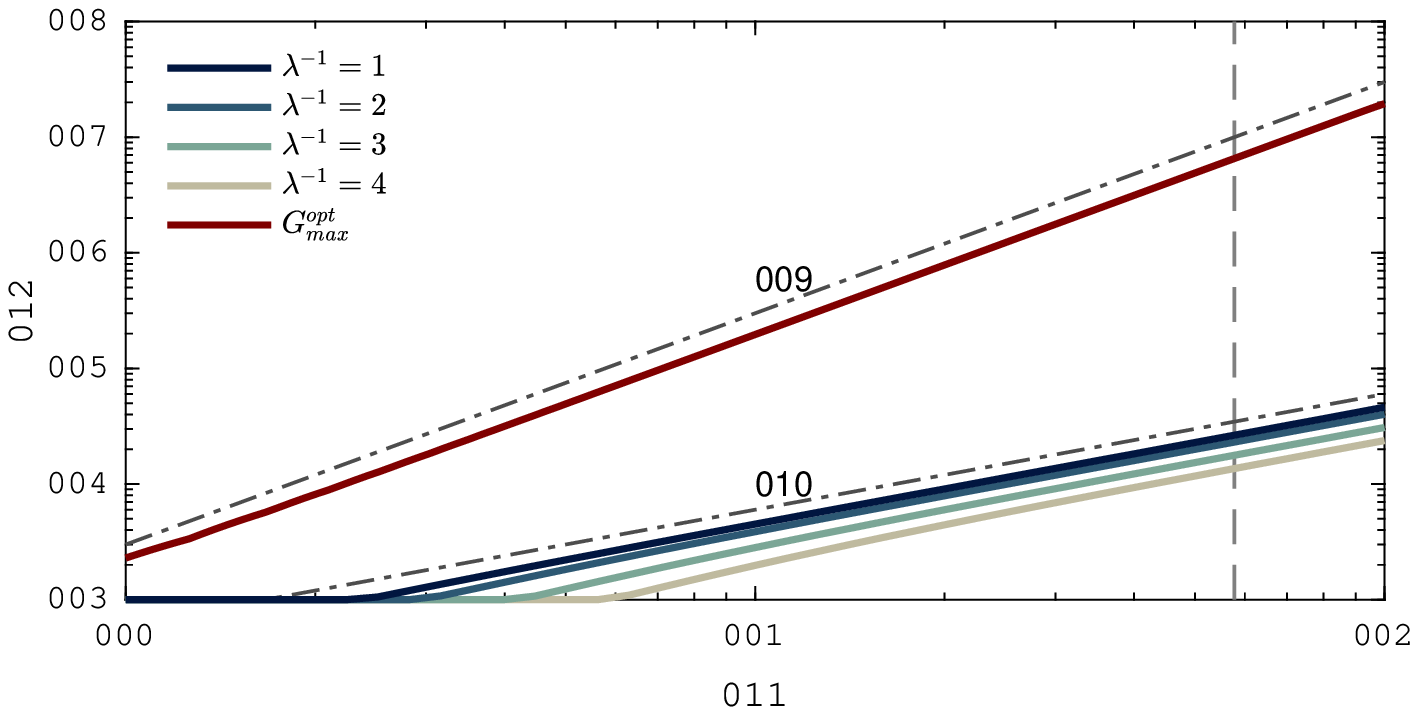}
    \caption{Scaling of $G^{mean}_{max}$ for the isotropic correlation for various correlation lengths along with that of $G^{opt}_{max}$. The gray guidelines show linear and quadratic Reynolds number scaling. Unlike $G^{opt}_{max}$, $G^{mean}_{max}$ with the isotropic correlation scales linearly with Reynolds number.}
    \label{fig:3D_reynolds_scaling}
\end{figure}
The Reynolds number scaling is shown in Figure~\ref{fig:3D_reynolds_scaling}. Unlike $G^{opt}_{max}$ or $G^{mean}_{max}$ at a particular wavenumber pair, $G^{mean}_{max}$ for an isotropic three-dimensional correlation scales nearly linearly with Reynolds number. This is a surprising result --- the three-dimensional $G^{mean}(t)$ is obtained in (\ref{eq:exenergy:3dgmean}) by integrating over $G^{mean}(t)$ at particular wavenumbers. However, critically, these single-wavenumber values for $G^{mean}$ peak at different times, as can be seen in Figure~\ref{fig:single_ab:lam_sweep}. The difference in the scaling means that the difference between $G^{mean}_{max}$ and $G^{opt}_{max}$ becomes larger with Reynolds number, i.e., $G^{opt}_{max}$ increasingly overpredicts the mean energy amplification with increasing Reynolds number.
\\

\subsubsection{Non-isotropic correlation}
As a generalization of the isotropic correlation investigated above, we next consider the ellipsoid
\begin{equation}
    \boldsymbol{C}^{vv}(\boldsymbol{x}_1,\boldsymbol{x}_2) = A\exp \left[-(\frac{\Delta x^2}{\lambda_x^2} + \frac{\Delta y^2}{\lambda_y^2} + \frac{\Delta z^2}{\lambda_z^2}) \right] \text{,}
\end{equation}
where $\lambda_x$, $\lambda_y$, and $\lambda_z$ are the correlation lengths in the streamwise, spanwise, and wall-normal directions, respectively. For example, this allows for the correlations to persist longer in $x$ than in $y$ or $z$, as may result from the advective nature of the flow \cite{He17}. With this extra freedom relative to the isotropic case, we may ask whether the Reynolds number scaling remains linear, as it is in that case, or becomes quadratic, as it is for $G^{opt}$. The answer depends on the correlation lengths chosen, but we find that if we fix $\lambda_x$ at some non-zero value and vary Reynolds number, the scaling is linear. 
\\

When discretized in $y$, the ellipsoid correlation becomes
\begin{equation}
    {\bf C}_{00}(\Delta x, \Delta z) = {\bf C}_{00}^{y}\exp \left[-(\frac{\Delta x^2}{\lambda_x^2} + \frac{\Delta z^2}{\lambda_z^2})\right] \text{,}
\end{equation}
where, again, ${\bf C}_{00}^{y}$ is the discretized $y$-dependant part. Upon taking the Fourier transform, the correlation in wavenumber space is
\begin{equation}
    \hat{\bf C}_{00}(\alpha,\beta) = A \exp \left[-\frac{1}{4}(\lambda_x^2 \alpha^2 + \lambda_z^2 \beta^2) \right]{\bf C}_{00}^{y} \text{.}
\end{equation}
Once again, we make use of (\ref{eq:exenergy:3dgmean}) to obtain
\begin{equation}
\begin{aligned}
    G^{mean}(t) = & \int_{-\infty}^{\infty} \Tr \{{\bf LM}_t(\alpha,\beta){\bf C}_{00}^{y}{\bf M}_t^*(\alpha,\beta){\bf L}^*\} \exp \left[-\frac{1}{4}(\lambda_x^2 \alpha^2 + \lambda_z^2 \beta^2) \right]{\bf C}_{00}^{y}  d\alpha \ d\beta \\
    & \bigg/ \int_{-\infty}^{\infty} \Tr \{{\bf LC}_{00}^{y}{\bf L}\} \exp \left[-\frac{1}{4}(\lambda_x^2 \alpha^2 + \lambda_z^2 \beta^2) \right]  d\alpha \ d\beta \text{.}    
\end{aligned}
\end{equation}
\\

\begin{figure} [!hbt]
    \centering
    \input{Figures_2/ellipsoid_maxall_crit.tex}
    \includegraphics{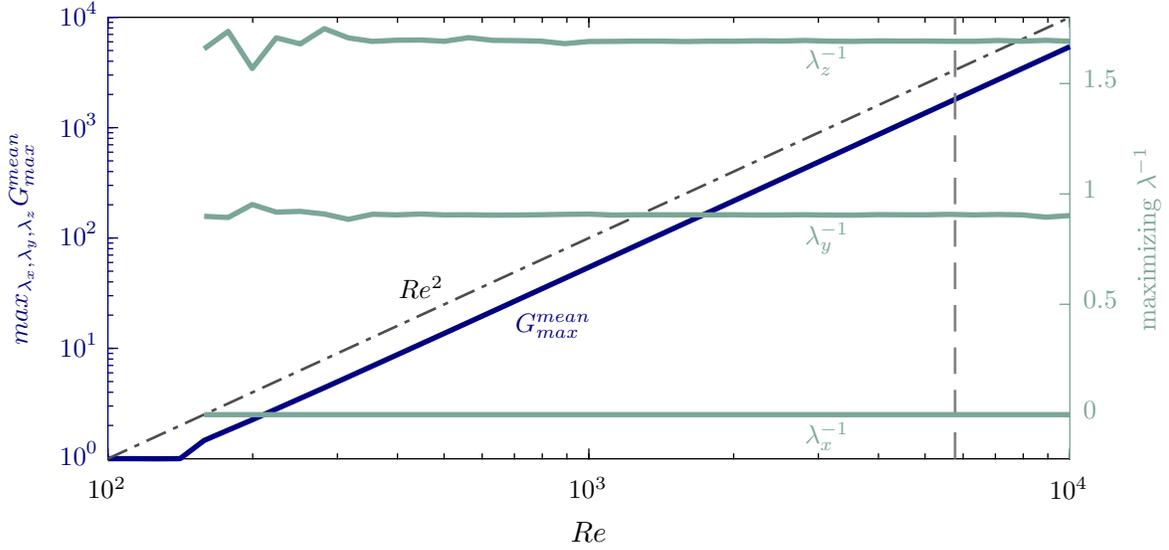}
    \caption{$G^{mean}_{max}$ optimized over the three correlation lengths (left axis) and the optimal correlation lengths (right axis) vs. $Re$. The scaling of $G^{mean}_{max}$ is quadratic here, and the optimal correlations do not change substantially with $Re$.}
    \label{fig:ellipsoid_maxall}
\end{figure}
The optimal correlation lengths and the resulting maximum value of $G^{opt}_{max}$ are shown in Figure \ref{fig:ellipsoid_maxall}. Plotted on the left axis in blue, $G^{mean}_{max}$ scales quadratically when the correlation lengths are optimized. At $Re = 1000$, $G^{mean}{max} = 54$, which is $27\%$ of $G^{opt}_{max}$ at the same Reynolds number. The optimal correlation lengths, plotted on the right axis in green, do not vary significantly with Reynolds number. The optimal $\lambda_y$ is close to the channel half-height, which is consistent with the most amplified wavenumbers shown in Figure~\ref{fig:single_ab:lam_sweep}. The maximizing $\lambda_x^{-1}$ is zero. This means that the disturbances are infinitely correlated in $x$, implying that their energy is concentrated at $\alpha = 0$. This wavenumber is known to produce the most growth (see, e.g., Figures \ref{fig:single_ab:abcontour} and \ref{fig:TG:Gmax}), so it is not surprising that the optimal $\lambda_x^{-1}$ is zero. The characteristic $\beta$ in the initial correlation is $2\lambda_z^{-1}$, so in light of Figure~\ref{fig:single_ab:abcontour}, the maximizing $\lambda_z^{-1}$ is not a surprise. The optimal correlation lengths are not reported in Figure~\ref{fig:ellipsoid_maxall} when $G^{mean}_{max} = 1$; this only occurs if $G^{mean}(t)$ peaks at the initial time $t = 0$, in which case any set of correlation lengths will produce the same result.
\\

\begin{figure} [!hbt]
    \centering
    \input{Figures_2/ellipsoid_scaling_crit.tex}
    \includegraphics{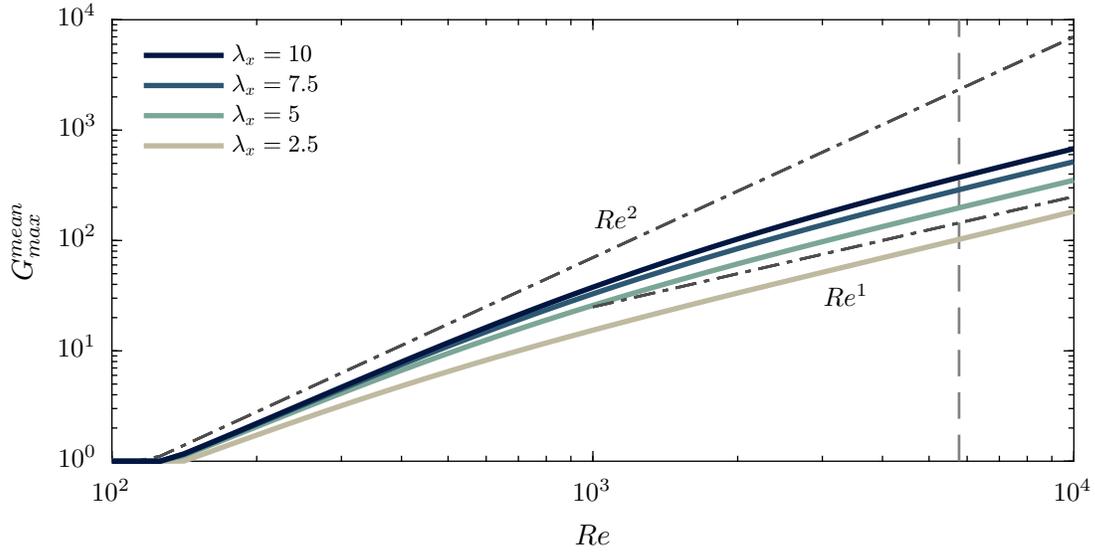}
    \caption{Scaling of $G^{mean}_{max}$ with $Re$ for $[\lambda_y^{-1},\lambda_z^{-1}] = [1,1.7]$ and various $\lambda_x$. The scaling is initially quadratic but becomes linear at higher $Re$. Longer correlations in $x$ remain quadratic to higher $Re$.}
    \label{fig:ellipsoid_scaling}
\end{figure}
The correlation will, in reality, be finite in $x$, and this can affect the scaling. Figure~\ref{fig:ellipsoid_scaling} shows $G^{mean}_{max}$ vs. Reynolds number for constant finite choices of $\lambda_x$, $\lambda_y$, and $\lambda_z$. The values of $\lambda_x$ are shown and $[\lambda_y^{-1},\lambda_z^{-1}] = [1,1.7]$, which are near their optimal values, as shown in Figure~\ref{fig:ellipsoid_maxall}. For small Reynolds numbers, the scaling is nearly quadratic, matching that of the infinitely correlated disturbances, but as the Reynolds number increases, the scaling once again becomes linear. The Reynolds number at which the scaling changes depends on $\lambda_x$, with longer correlations remaining quadratic up to higher Reynolds numbers. 
\\

\section{Estimating the PDF} \label{sec:Estimate_PDF}
As Figures \ref{fig:single_ab:lam_sweep} and \ref{fig:3d_lamsweep} show, the expected value of the energy may be orders of magnitude smaller than $G^{opt}_{max}$ for certain reasonable incoming correlations. With a formula for the expected energy (\ref{Gmean}), one may still ask whether large gains are possible or negligibly unlikely. In this section, we discuss methods to estimate the probability density function (PDF) of the energy. Whereas the mean energy depends only on the initial correlations, the PDF depends on the entire distribution of the incoming disturbances. We first describe a basic Monte Carlo approach for estimating the PDF and apply this to two candidate distributions for the incoming disturbances, noticing that the PDF of the energy drops nearly exponentially. In the two subsequent subsections, two distributions for the initial disturbances are considered, a multivariate Gaussian and a transformation of the uniform distribution on the $N$-sphere. For these distributions, it is possible to analytically approximate the PDF of the energy using the exactly calculable moments of the energy distribution. With an accurate estimate of the PDF, we can calculate percentile curves for the trajectories.
\\

We denote the probability density function of a random variable $X$ as $f_X(x) \equiv \lim_{dx \to 0} Pr\{ X \in [x,x+dx] \} / dx$, where $Pr\{ \cdot \}$ denotes probability. If $X \in \mathbb{R}^N$ is a vector, then the PDF is defined 
\begin{equation}
    f_X({\bf x}) = f_X(x_1, \dots , x_N) \equiv \lim_{dx_1 \to 0, \dots, dx_N \to 0} \frac{Pr\{ X_1 \in [x_1,x_1 + dx_1] , \dots , X_N \in [x_N,x_N + dx_N]\} } {dx_1\cdots dx_N} \text{.}
\end{equation}
The incoming disturbances follow some distribution
\begin{equation}
    {\bf q}(0) \sim f_{{\bf q}(0)}({\bf q}_0) \text{,}
\end{equation}
and this implies a distribution
\begin{equation}
        {\bf q}(t) \sim f_{{\bf q}(t)}({\bf q}_t)
\end{equation}
for the disturbances ${\bf q}(t) = {\bf M}_t{\bf q}(0)$ at time $t$. These distributions are the most descriptive statistical information about the disturbances; any statistic of the disturbances is implied by the full distribution. For example, the correlation matrix ${\bf C}_{00}$ for the initial disturbances is implied by the distribution of the initial disturbance $f_{{\bf q}(0)}$. The converse, however, is not true --- there are many distributions with the same correlation matrix. In the last section, we showed that $G^{mean}$ only depends on the correlation matrix, so there was no need to consider the form of the underlying distribution. However, to calculate the PDF of the energy of the disturbance at some time,
\begin{equation}
    e(t) \sim f_{e(t)}(e_t) \text{,}
\end{equation}
the full distribution of initial disturbances is needed. 
\\

\subsection{Monte Carlo}
\begin{figure}
    \centering
    \input{Figures_2/empirical_PDF.tex}
    \includegraphics{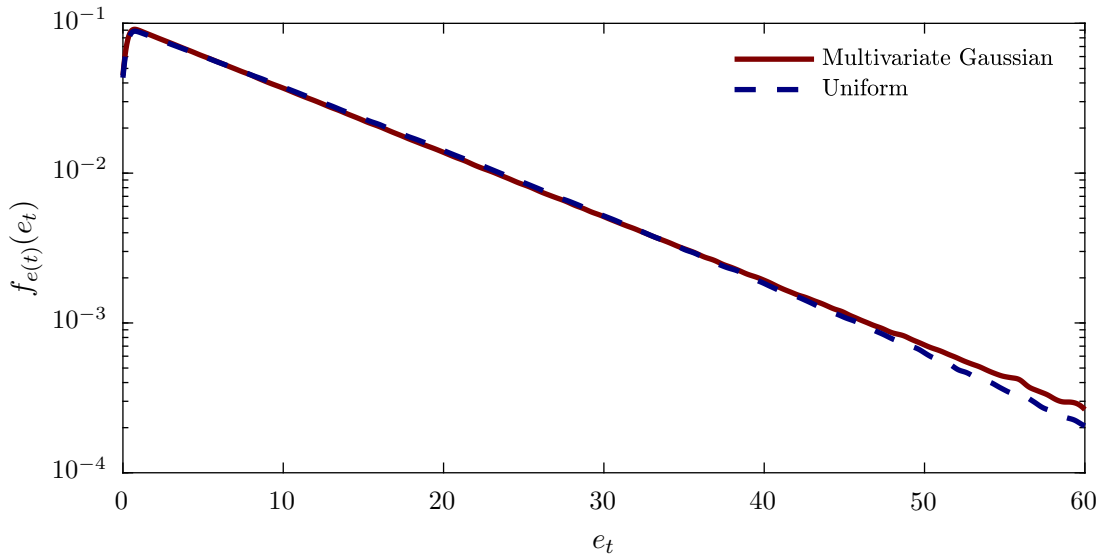}
    \caption{Empirical PDF of energy at $t = 100$ resulting from initial disturbances distributed as a transformation of the uniform distribution and a multivariate Gaussian, both with the same correlation with correlation length $\lambda^{-1} = 5$. Note the similarity in the two PDFs and the near-exponential decay.}
    \label{fig:monte_carlo:both}
\end{figure}

With a means of sampling initial disturbances from $f_{{\bf q}(0)}({\bf q}_0)$, samples of the growing disturbances can be generated by multiplying the initial ones by ${\bf M}_t$, and samples of $e(t)$ are finally obtained by computing their norm. An estimation of the PDF can be obtained using standard methods, such as $\texttt{ksdensity}$ in Matlab. Figure \ref{fig:monte_carlo:both} shows the empirical PDF resulting from performing this Monte Carlo with two different distributions of the initial disturbances (described later), both with the same correlation. The Monte Carlo is performed at $t = 100$, $Re = 1000$, $\alpha = 0$, $\beta = 2$ with a correlation of the form (\ref{eq:singleab:correlation_form}) and correlation length $\lambda^{-1} = 5$. The distributions are described in detail in the following subsections. Two observations are apparent. First, both distributions result in a very similar PDF for the energy of the amplified disturbance. That the PDFs are similar indicates that while the PDF of the energy is a function of the full distribution of incoming disturbances, reasonable distributions of initial disturbances with the same correlation will give similar PDFs for the energy. Second, the PDFs decay nearly exponentially. The exponential decay indicates that it is very unlikely that the energy of an amplified disturbance substantially exceeds $G^{mean}(t)$. The exponential decay also allows for accurate a priori approximation of the PDF, which we discuss in the following subsections for the two distributions of initial disturbances.
\\

\subsection{Multivariate Gaussian}
First, we assume the initial disturbances follow the multivariate Gaussian with mean $\boldsymbol{0}$ and correlation ${\bf C}_{00}$,
\begin{equation}
        {\bf q}(0) \sim \mathcal{N}({\bf 0},{\bf C}_{00}) \text{.}
\end{equation}
Also, we assume the correlation to have unit trace, so $G^{mean}(t) = \mathbb{E}[e({\bf q}(t))]$. More explicitly, this distribution, as represented by its PDF, is
\begin{equation}
    f_{{\bf q}(0)}({\bf q}_0) = \frac{\exp[-\frac{1}{2}{\bf q}_0^* {\bf C}_{00}^{-1} {\bf q}_0]} {\sqrt{(2\pi)^N | {\bf C}_{00}|}} \text{.}
\end{equation}
If ${\bf C}_{00}$ is rank-deficient, the inverse ${\bf C}_{00}^{-1}$ and determinant $|{\bf C}_{00}|$ are modified to the pseudoinverse and pseudodeterminant.
\\

Any linear function of ${\bf q}(0)$ also follows a multivariate Gaussian distribution \cite{Tong90}, so the disturbance ${\bf q}(t) = {\bf M}_t {\bf q}(0)$ some time in the future is distributed as
\begin{equation}
        {\bf q}(t) \sim \mathcal{N}({\bf 0},{\bf C}_{tt}) \text{,}
\end{equation}
where the correlation ${\bf C}_{tt} = {\bf M}_t {\bf C}_{00} {\bf M}_t^*$ comes from (\ref{Correlation_out}).
The energy is $e(t) = \|{\bf q}(t) \|^2$ and we seek to estimate its PDF $f_{e(t)}(e_t)$. This distribution is one of a well-studied class --- quadratic forms in multivariate Gaussian variables. The moments of these distributions are known \cite{Mathai92}, and for the case at hand, the $r$-th moment $\mu_r$ can be calculated recursively as
\begin{subequations} \label{Moments_gaussian}

\begin{equation} 
    \mu_r \equiv \mathbb{E}[e^r] = \sum_{k = 0}^{r-1}\binom{r-1}{k}g^{(r-1-k)}\mu_{k} \text{,}
\end{equation}

\begin{equation}
    \text{where} \quad g^{(k)} = 2^kk! \Tr\{  \left( {\bf LC}_{tt} {\bf L}^* \right)^{k+1} \} \text{,}
\end{equation}
\end{subequations}
and $\mu_0 = 1$. Note that $\mu_1 = G^{mean}$, i.e., the first moment recovers the expected energy.
\\

\begin{figure}
    \centering
    \input{Figures_2/disp_approx.tex}
    \includegraphics{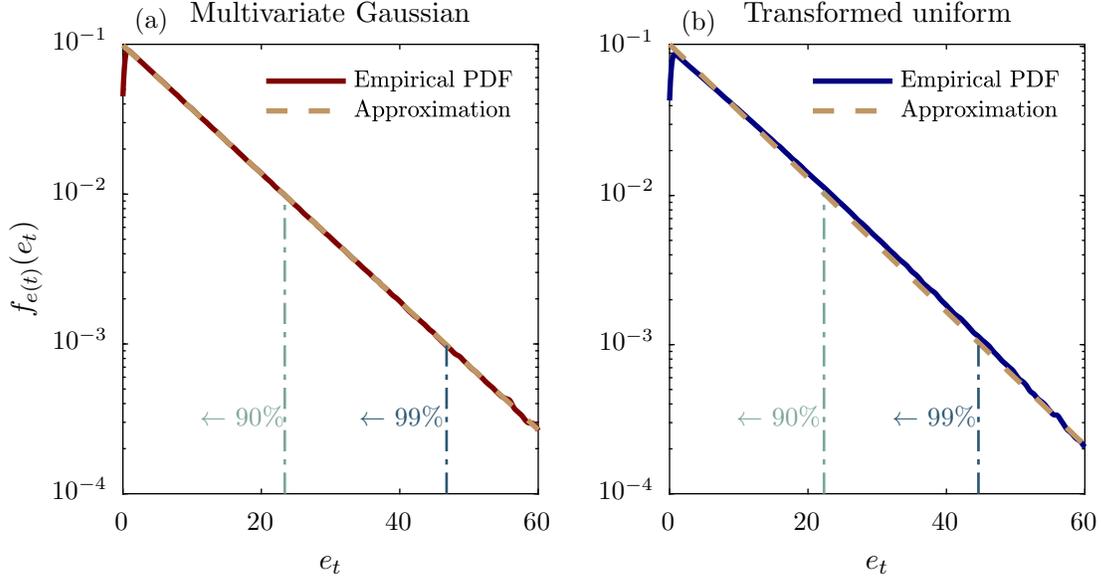}
    \caption{Empirical PDF for the disturbance energy and the approximation thereof. The approximation is the exponential distribution with the same fourth moment as the true distribution and is calculated without the Monte Carlo.}
    \label{fig:pdf_approx_both}
\end{figure}

Our goal is to estimate the right tail of the PDF of the energy in order to approximate the probability of exceeding a particular energy. As shown in Figure \ref{fig:monte_carlo:both}, the right tail of the empirical PDF displays nearly exponential decay. Therefore we assume its form to be 
\begin{equation} \label{eq:PDF:exp_pdf_def}
    f_{e(t)}(e_t) \approx \gamma\exp[-\gamma e_t] \text{.}
\end{equation}
To find the decay rate, we find $\gamma$ such that the $r$-th moment of the exponential ansatz matches the $r^{\text{th}}$ moment of the true distribution, given in (\ref{Moments_gaussian}). The true distribution, estimated via the Monte Carlo, is near-exponential for high energies but not low ones, so to find the correct exponential parameter we equate a relatively high moment, since this weights the high-energy tail of the distribution heavily. Denoting the moment equated as $r$, the $r$-th moment of the exponential distribution (\ref{eq:PDF:exp_pdf_def})  is $\mu_r = \frac{r!}{\gamma^r}$. Equating this to the true $r$-th moment given in (\ref{Moments_gaussian}) and solving for the exponential decay rate gives
\begin{equation}
    \gamma = \big(\frac{r!}{\mu^r}\big)^{1/r} \text{.}
\end{equation}
This is an analytical approximation; the only role of the previous Monte Carlo was to suggest the exponential form (\ref{eq:PDF:exp_pdf_def}). Figure \ref{fig:pdf_approx_both}(a) shows this approximation strategy using $r = 4$. The confidence bounds are derived by integrating the approximate PDF. The Reynolds number, wavenumbers, and correlation length are $Re = 1000$, $\alpha = 0$, $\beta = 2$, $\lambda^{-1} = 5$. Out of the $10^7$ trajectories used to generate the empirical distribution, $99.005 \%$ were below the $99\%$ confidence bound.

\subsection{Transformation of a uniform distribution on the $N$-sphere}
Second, we assume that the disturbances are distributed as some transformation of a uniform distribution on the surface of the $N$-sphere,
\begin{equation} \label{eq:uni:transformed_uni}
        {\bf q}(0) \sim {\bf \Psi}{\bf u} \text{,}
\end{equation}
where $\|u\| = 1 $ with probability $1$, and ${\bf u} \sim {\bf Ru}$ for any rotation matrix ${\bf R}$. Samples from this distribution can be easily generated by normalizing samples from an i.i.d. multivariate Gaussian to be the same radius $R$, ${\bf u} = R\frac{{\bf x}}{\| {\bf x}\|}$ with ${\bf x} \sim \mathcal{N}(0,{\bf I})$. The transformation ${\bf \Psi}$ can be chosen so that the initial disturbances have some desired correlation ${\bf C}_{00}$. By choosing ${\bf \Psi}$ such that
\begin{equation}
    {\bf \Psi \Psi}^* = N{\bf C}_{00}  \text{,}
\end{equation}
e.g., setting it to the Cholesky decomposition of $N{\bf C}_{00}$, where $N$ is the dimension of the state, the correlation matrix of the initial disturbances is 
\begin{equation}
    \mathbb{E}[{\bf \Psi uu}^*{\bf \Psi}^*] = \frac{1}{N}{\bf \Psi \Psi}^* = {\bf C}_{00} \text{.}
\end{equation}
The first equality holds because $\mathbb{E}[{\bf u u}^*] = \frac{1}{N}{\bf I}$. The disturbance energy some time later is given by $e(t) = \|{\bf q}(t) \|^2 = \|{\bf T u}\|^2$. The final equality expresses the energy as the square norm of the uniform distribution acted on by a matrix ${\bf T} = {\bf M}_t{\bf \Psi}$.
\\

The moments for this transformation of the uniform distribution were derived by von Neumann \cite{von_Neumann41}. The result is \cite{Kargan10}
\begin{equation}
    \mathbb{E}[e^k] = \frac{2^kk!}{N(N+2)(N+4) \dots (N+2k-2)}\xi_k \text{,}
\end{equation}
where the $\xi$ are defined as the power series coefficients of 
\begin{equation}
    1 + \xi_1z + \xi_2 z^2 + \xi_3 z^3 + \dots = \exp[\zeta_1z + \zeta_2 z^2 + \zeta_3 z^3 + \dots] \text{,}
\end{equation}
and $\zeta_j$ is defined in terms of the singular values $\sigma_i({\bf LT})$, as
\begin{equation}
    \zeta_j = \frac{1}{2j}\sum_{i = 1}^N\sigma_i^{2i} \text{.}
\end{equation}
Using these moments, we can approximate the true PDF using the technique described in the previous subsection. Figure~\ref{fig:pdf_approx_both} shows the result along with confidence bounds derived by integrating the approximated PDF. Just as before, the Reynolds number, wavenumbers, and correlation length are $Re = 1000$, $\alpha = 0$, $\beta = 2$, $\lambda^{-1} = 5$. Of the $10^7$ trajectories used to calculate the empirical distribution, $98.981\%$ were less than the $99\%$ confidence bound. 
\\

\section{Conclusions} \label{sec:Conclusions}
Standard transient growth analyses are based on the maximum growth experienced by any initial disturbance. While this is a useful upper bound on linear growth, it can vastly overpredict the growth experienced by real disturbances. We have developed a statistical framework to explore the space of real disturbances and quantify their growth. We demonstrated the framework and its ability to extract insights on Poisseuille flow. 
\\

\begin{figure}
    \centering
    \input{Figures_2/trajs_levels2.tex}
    \includegraphics{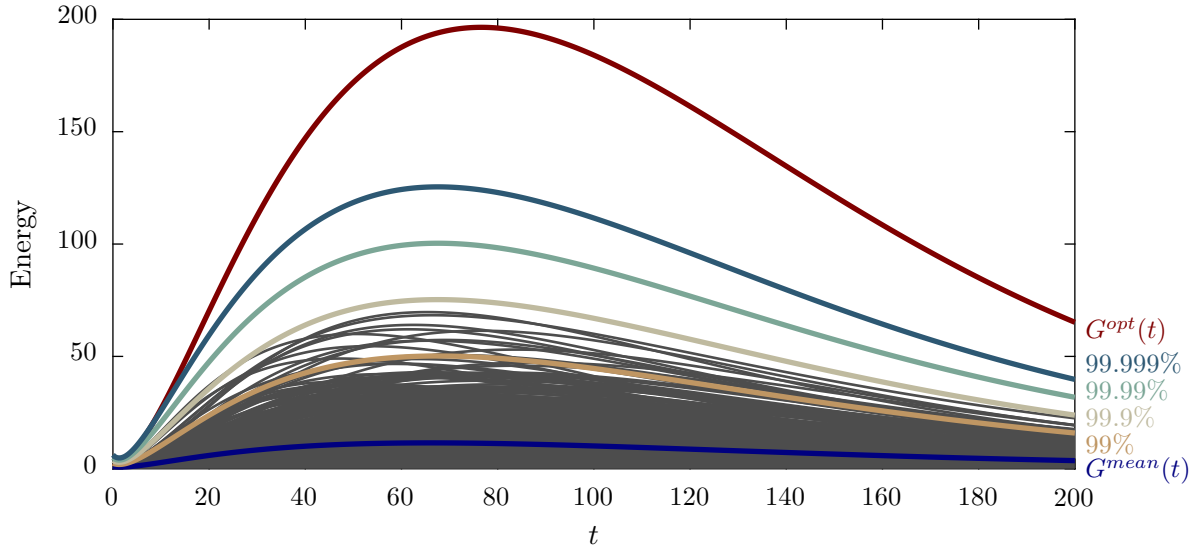}
    \caption{The one-thousand trajectories in Figure~\ref{fig:TG:trajs} overlayed with the calculated mean and percentile curves. The Reynolds number is $1000$, the wavenumbers are $\alpha = 0$, $\beta = 2$, and $\lambda^{-1} = 5$ with the correlation given in (\ref{eq:singleab:correlation_form}).  The mean drastically undershoots the optimum, and the energy is exponentially distributed, so it is unlikely for a disturbance to grow by near $G^{opt}$.}
    \label{fig:PDF:trajs_levels}
\end{figure}
The framework can be summarized using Figure~\ref{fig:PDF:trajs_levels}. $G^{opt}$, the quantity used in the literature to quantify transient growth, far overshoots all one thousand random trajectories, which are the same ones shown in Figure~\ref{fig:TG:trajs}. $G^{mean}(t)$, gives the mean energy divided by the expected initial energy, i.e., the mean energy amplification. It is a function of the correlation matrix of the initial disturbances but does not depend explicitly on the particular form of the distribution of initial disturbances. As Figure~\ref{fig:PDF:trajs_levels} shows, $G^{mean}(t)$ can be significantly lower than $G^{opt}(t)$. For three-dimensional disturbances, the three-dimensional correlation matrix determines the average energy at each wavenumber. With a realistic correlation matrix, suboptimal wavenumbers will account for a significant portion of the energy, and this, coupled with the fact that different wavenumbers peak at different times, further widens the gap between $G^{mean}$ and $G^{opt}$. The confidence bounds in Figure~\ref{fig:PDF:trajs_levels} give the energy levels that $p \%$ of the trajectories undershoot. The levels are calculated analytically by integrating the PDF of the energy, not by performing a Monte Carlo. The energy PDF cannot be calculated exactly for a general distribution of initial disturbances; however, because the energy PDF is nearly exponential, it can be approximated accurately.
\\

Applied to Poisseuille flow, this statistical view reveals a number of insights. For a single wavenumber pair, the correlation length in the wall-normal direction emerges as an important parameter in determining $G^{mean}$. For long correlation lengths, $G^{mean}$ is on the same order as $G^{opt}$ (as much as half for certain wavenumbers), but for short correlation lengths, $G^{mean}$ is orders of magnitude smaller than $G^{opt}$. The dependence of $G^{mean}$ on the streamwise and spanwise wavenumbers is different than that of $G^{opt}$: while the peak is still near $\alpha = 0$, $\beta = 2$, it is substantially narrower in $\alpha$. Three-dimensional disturbances contain energy at all wavenumber pairs, so the narrower peak of $G^{mean}$ in $\alpha$ leads to substantially less growth. At $Re = 1000$, with an isotropic correlation of correlation length $\lambda$, $G^{mean}_{max}$ is only $2 \%$ of $G^{opt}$ even at the most growth-promoting $\lambda$. Furthermore, for this form of the three-dimensional correlation, $G^{mean}_{max}$ grows near-linearly, while $G^{opt}_{max}$ grows quadratically, leading the latter to increasingly overpredict the mean energy amplification as the Reynolds number increases.
\\

The formulae derived depend on the correlation matrix of the initial disturbances. We are not aware of previous studies on the statistics of these disturbances in Poisseuille flow, but they are likely to be dependent on the source of disturbances. The figures reported, e.g., for the ratio of $G^{mean}/G^{opt}$, are not meant to be taken as quantitative predictions of what would be observed in an experiment. Rather, they are meant to show trends and to emphasize that, when various factors are accounted for, the growth of real disturbances can be significantly smaller than $G^{max}$. Though $G^{mean}$ depends on the full correlation matrix, the correlation length is a particularly important feature in determining $G^{mean}$. The observation that long correlations lead to more growth may serve as a practical guide when an accurate model for correlations is unavailable. 
\\

We have also described a statistical approach to determining the structures that emerge, in the form of modes from a particular POD problem. These structures also depend on the initial correlations but were observed only to differ slightly from the standard output modes of the evolution operator, which do not depend on the correlations. This indicates that the output modes do a good job of capturing the energy of the growing disturbances regardless of the correlation matrix. Nevertheless, they do not accurately predict the energy of each structure, whereas the POD eigenvalues do.
\\

We have discussed the statistical framework in the context of temporal stability, wherein an initial disturbance at a particular time is assumed, then evolved forward in time without further forcing to the linear dynamics. In spatial stability, a disturbance is introduced at a particular streamwise location, and its growth is then calculated as it evolves downstream as a function of the streamwise coordinate. Transient growth has been investigated in the context of spatial stability \cite{Hack17, Hanifi22}, and the framework developed in this paper applies equally to spatial stability by exchanging $t$ for $x$ and the linearized Navier-Stokes operator for a spatial evolution operator \cite{Towne15}. Spatial stability may be an excellent application for two reasons. First, applied to spatial stability, $G^{mean}(x_1)$ represents the ratio of turbulent intensities at $x = x_1$ and $x = 0$. Consider a laminar boundary layer excited at $x = 0$ by free-stream turbulence. The turbulent intensity at various points along the boundary layer is critical for determining where transition occurs, and its evolution in space is given by $G^{mean}(x)$. In this context, we have shown that the turbulent intensity depends not only on the input turbulent intensity, encapsulated by the diagonal terms of the initial correlation matrix, but on the off-diagonal terms as well. Second, the correlation matrix of the initial disturbances may be easier to model physically in spatial stability than in temporal stability. In this case, incoming turbulence is often the source of the disturbances and turbulent correlations have been studied thoroughly. In the example of a boundary layer, the free-stream turbulence that excites the leading edge may be modeled by the von K\'arm\'an spectrum \cite{vonKarman48}, which implies the correlations within the initial disturbances. Also, with such a model for the initial correlation, the POD modes of the space-evolved correlation may provide a better basis for the structures that arise downstream than the output modes of the matrix exponential.
\\

\begin{appendices}

% Proof of gmean equality with mean growth for separable density
\section{PDF separability} \label{App:first_appendix}
If the distribution of the initial disturbance is separable in radius and direction,
\begin{equation}
    {\bf q}(0) \sim f_{{\bf q}(0)}({\bf q}_0) = R(\|{\bf q}_0\|)\Theta \left(\frac{{\bf q}_0}{\|{\bf q}_0\|} \right) \text{,}
\end{equation}
the ratio of expected energies $G^{mean}$ is equal to the expected ratio of energies,
\begin{equation} \label{To_prove}
    \frac{\mathbb{E}[e({\bf q}(t))]}{\mathbb{E}[e({\bf q}(0))]} = \mathbb{E} \left[\frac{e({\bf q}(t))}{e({\bf q}(0))} \right] \text{.}
\end{equation}
This can be shown as follows. Assume, without loss of generality, that $R$ and $\Theta$ are scaled such that they both integrate to unity (with the appropriate measure). Defining $r^2 \equiv {\bf q}_0^*{\bf W}{\bf q}_0 = \|{\bf Lq}_0\|^2$ and $\boldsymbol{\theta} \equiv {\bf q}_0/r$, we write the LHS of (\ref{To_prove}) in terms of the distribution,
\begin{align}
    \frac{\mathbb{E}[e({\bf q}(t))]}{\mathbb{E}[e({\bf q}(0))]} & = \int_{0}^{\infty} dr \int_{\|\boldsymbol{\theta}\| = 1} R(r) \Theta(\boldsymbol{\theta})\| {\bf M}_t{\bf L}r\boldsymbol{\theta} \|^2 d\sigma(\boldsymbol{\theta}) \bigg/ \int_{0}^{\infty} dr \int_{\|\boldsymbol{\theta}\| = 1} R(r) \Theta(\boldsymbol{\theta}) r^2 d\sigma(\boldsymbol{\theta}) \\
     & = \int_{0}^{\infty} R(r) r^2 dr \int_{\|\boldsymbol{\theta}\| = 1} \Theta(\boldsymbol{\theta})\| {\bf M}_t{\bf L}\boldsymbol{\theta} \|^2 d\sigma(\boldsymbol{\theta}) \bigg/ \int_{0}^{\infty}R(r) r^2  dr \int_{\|\boldsymbol{\theta}\| = 1} \Theta(\boldsymbol{\theta})d\sigma(\boldsymbol{\theta}) \\
    & = \int_{\|\boldsymbol{\theta}\| = 1} \Theta(\boldsymbol{\theta}) \| {\bf M}_t{\bf L}\boldsymbol{\theta} \|^2 d\sigma(\boldsymbol{\theta}) \text{.}
\end{align}
Above, $\sigma(\boldsymbol{\theta})$ is the spherical measure. The RHS of (\ref{To_prove}) can be written,
\begin{align}
\mathbb{E}\left[\frac{e({\bf q}(t))}{e({\bf q}(0))}\right] & = \int_{0}^{\infty} dr \int_{\|\boldsymbol{\theta}\| = 1} R(r) \Theta(\boldsymbol{\theta}) \frac{\| {\bf M}_t{\bf L}r\boldsymbol{\theta} \|^2}{r^2} d\sigma(\boldsymbol{\theta}) \\
& = \int_{0}^{\infty}R(r) dr \int_{\|\boldsymbol{\theta}\| = 1}   \Theta(\boldsymbol{\theta}) \| {\bf M}_t{\bf L}\boldsymbol{\theta} \|^2 d\sigma(\boldsymbol{\theta}) \\
& = \int_{\|\boldsymbol{\theta}\| = 1} \Theta(\boldsymbol{\theta})\| {\bf M}_t{\bf L}\boldsymbol{\theta} \|^2 d\sigma(\boldsymbol{\theta}) \text{,}
\end{align}
so the two are equal and (\ref{To_prove}) holds.

\end{appendices}

%%%%%%%%%%%%%%%%%%%%%%%%%%%%%%%%%%%%%%%%%%%%%%%%%%%%%%%%%%%%%%%%%%%%%%%%%%%%%%%
% -----------------------------------------------------------------------------
% -- Backmatter 
% -----------------------------------------------------------------------------

% Acknowledgments

% Bibliography
\bibliographystyle{abbrv}
\bibliography{bibliography}

\begin{thebibliography}{10}

\bibitem{Hiroyuki01}
H.~Abe, H.~Kawamura, and Y.~Matsuo.
\newblock {Direct Numerical Simulation of a Fully Developed Turbulent Channel
  Flow With Respect to the Reynolds Number Dependence }.
\newblock {\em J. Fluids Eng.}, 123(2):382--393, 02 2001.

\bibitem{Butler92}
K.~M. Butler and B.~F. Farrell.
\newblock Three‐dimensional optimal perturbations in viscous shear flow.
\newblock {\em . Phys. Fluids A}, 4(8):1637--1650, 1992.

\bibitem{Farrell93}
B.~F. Farrell and P.~J. Ioannou.
\newblock Stochastic forcing of the linearized navier–stokes equations.
\newblock {\em . Phys. Fluids A}, 5(11):2600–2609, Nov 1993.

\bibitem{Hanifi22}
T.~C.~L. Fava, B.~A. Lobo, A.~P. Schaffarczyk, M.~Breuer, A.~Hanifi, and
  D.~Henningson.
\newblock On the stability and transition to turbulence of the flow over a
  wind-turbine airfoil under varying free-stream turbulence intensity.
\newblock 12th International Symposium on Turbulence and Shear Flow Phenomena,
  2022.

\bibitem{Frame22}
P.~Frame and A.~Towne.
\newblock Space-time pod and the hankel matrix.
\newblock 2022.

\bibitem{Gustavsson86}
L.~H. Gustavsson.
\newblock Excitation of direct resonances in plane poiseuille flow.
\newblock {\em Stud. Appl. Math.}, 75(3):227--248, 1986.

\bibitem{Gustavsson91}
L.~H. Gustavsson.
\newblock Energy growth of three-dimensional disturbances in plane poiseuille
  flow.
\newblock {\em J. Fluid Mech.}, 224:241–260, 1991.

\bibitem{Hack17}
M.~J.~P. Hack and P.~Moin.
\newblock Algebraic disturbance growth by interaction of orr and lift-up
  mechanisms.
\newblock {\em J. Fluid Mech.}, 829:112–126, 2017.

\bibitem{Hanifi96}
A.~Hanifi, P.~J. Schmid, and D.~S. Henningson.
\newblock Transient growth in compressible boundary layer flow.
\newblock {\em Phys. Fluids}, 8(3):826--837, 1996.

\bibitem{He17}
G.~He, G.~Jin, and Y.~Yang.
\newblock Space-time correlations and dynamic coupling in turbulent flows.
\newblock {\em Annu. Rev. Fluid Mech.}, 49(1):51--70, 2017.

\bibitem{Herbert77}
T.~Herbert.
\newblock {\em Die neutrale Fläche der ebenen Poiseuille-Strömung}.
\newblock habilitation, Uni Stuttgart, 1977.

\bibitem{Kargan10}
V.~Kargin.
\newblock Products of random matrices: Dimension and growth in norm.
\newblock {\em Ann. Appl. Probab.}, 20(3):890--906, 2010.

\bibitem{vonKarman48}
T.~V. Karman.
\newblock Progress in the statistical theory of turbulence.
\newblock {\em Proc. Natl. Acad. Sci. U.S.A.}, 34(11):530--539, 1948.

\bibitem{Kerswell18}
R.~Kerswell.
\newblock Nonlinear nonmodal stability theory.
\newblock {\em Annu. Rev. Fluid Mech.}, 50(1):319--345, 2018.

\bibitem{Khintchine34}
A.~Khintchine.
\newblock Korrelationstheorie der stationären stochastischen prozesse.
\newblock {\em Math. Ann.}, 109:604--615, 1934.

\bibitem{Lumley67}
J.~L. LUMLEY.
\newblock The structure of inhomogeneous turbulent flows.
\newblock {\em Atmospheric Turbulence and Radio Wave Propagation}, 1967.

\bibitem{Lumley70}
J.~L. Lumley.
\newblock Stochastic tools in turbulence.
\newblock 1970.

\bibitem{Malkus56}
W.~V.~R. Malkus.
\newblock Outline of a theory of turbulent shear flow.
\newblock {\em J. Fluid Mech.}, 1(5):521–539, 1956.

\bibitem{Markeviciute22}
V.~Markeviciute.
\newblock Statistical stability and fast transient growth in wall-bounded
  turbulence.
\newblock 2022.

\bibitem{Mathai92}
A.~Mathai and S.~Provost.
\newblock Quadratic forms in random variables.
\newblock 01 1992.

\bibitem{McKeon17}
B.~J. McKeon.
\newblock The engine behind (wall) turbulence: perspectives on scale
  interactions.
\newblock {\em J. Fluid Mech.}, 817:P1, 2017.

\bibitem{Pringle12}
C.~C. Pringle, A.~P. Willis, and R.~R. Kerswell.
\newblock Minimal seeds for shear flow turbulence: using nonlinear transient
  growth to touch the edge of chaos.
\newblock {\em J. Fluid Mech.}, 702:415–443, 2012.

\bibitem{Pringle10}
C.~C.~T. Pringle and R.~R. Kerswell.
\newblock Using nonlinear transient growth to construct the minimal seed for
  shear flow turbulence.
\newblock {\em Phys. Rev. Lett.}, 105:154502, Oct 2010.

\bibitem{Reddy93}
S.~C. Reddy and D.~S. Henningson.
\newblock Energy growth in viscous channel flows.
\newblock {\em J. Fluid Mech.}, 252:209–238, 1993.

\bibitem{Reynolds1883}
O.~Reynolds.
\newblock An experimental investigation of the circumstances which determine
  whether the motion of water shall be direct or sinuous, and of the law of
  resistance in parallel channels.
\newblock {\em Philosophical Transactions of the Royal Society of London},
  174:935--982, 1883.

\bibitem{Rowley17}
C.~W. Rowley and S.~T. Dawson.
\newblock Model reduction for flow analysis and control.
\newblock {\em Annu. Rev. Fluid Mech.}, 49(1):387--417, 2017.

\bibitem{SH}
P.~Schmid and D.~Henningson.
\newblock {\em Stability and Transition in Shear Flows}.
\newblock Springer, 2001.

\bibitem{Schmid07}
P.~J. Schmid.
\newblock Nonmodal stability theory.
\newblock {\em Annu. Rev. Fluid Mech.}, 39(1):129--162, 2007.

\bibitem{SH94}
P.~J. Schmid and D.~S. Henningson.
\newblock Optimal energy density growth in hagen–poiseuille flow.
\newblock {\em J. Fluid Mech.}, 277:197–225, 1994.

\bibitem{Sirovich87}
L.~Sirovich.
\newblock Turbulence and the dynamics of coherent structures. i - coherent
  structures. ii - symmetries and transformations. iii - dynamics and scaling.
\newblock {\em Quart. Appl. Math.}, 45, 10 1987.

\bibitem{Taira18}
K.~Taira, S.~L. Brunton, S.~T.~M. Dawson, C.~W. Rowley, T.~Colonius, B.~J.
  McKeon, O.~T. Schmidt, S.~Gordeyev, V.~Theofilis, and L.~S. Ukeiley.
\newblock Modal analysis of fluid flows: An overview.
\newblock {\em AIAA J.}, 55(12):4013--4041, 2017.

\bibitem{tillmark_alfredsson_1992}
N.~Tillmark and P.~H. Alfredsson.
\newblock Experiments on transition in plane couette flow.
\newblock {\em J. Fluid Mech.}, 235:89–102, 1992.

\bibitem{Tong90}
Y.~L. Tong.
\newblock {\em The Multivariate Normal Distribution}.
\newblock Springer, New York, NY, 2011.

\bibitem{Towne15}
A.~Towne and T.~Colonius.
\newblock One-way spatial integration of hyperbolic equations.
\newblock {\em Journal of Computational Physics}, 300:844--861, 2015.

\bibitem{Towne18}
A.~Towne, O.~T. Schmidt, and T.~Colonius.
\newblock Spectral proper orthogonal decomposition and its relationship to
  dynamic mode decomposition and resolvent analysis.
\newblock {\em J. Fluid Mech.}, 847:821–867, May 2018.

\bibitem{Trefethen05}
L.~N. Trefethen and M.~Embree.
\newblock {\em Spectra and Pseudospectra}.
\newblock Princeton University Press, Princeton, 2005.

\bibitem{Trefethen91}
L.~N. Trefethen, A.~E. Trefethen, S.~C. Reddy, and T.~A. Driscoll.
\newblock Hydrodynamic stability without eigenvalues.
\newblock {\em Science}, 261(5121):578--584, 1993.

\bibitem{von_Neumann41}
J.~von Neumann.
\newblock Distribution of the ratio of the mean square successive difference to
  the variance.
\newblock {\em The Annals of Mathematical Statistics}, 12(4):367--395, 1941.

\bibitem{Wiener30}
N.~Wiener.
\newblock {Generalized harmonic analysis}.
\newblock {\em Acta Math.}, 55(none):117 -- 258, 1930.

\end{thebibliography}

\end{document}